\documentclass[a4paper,11pt]{article}
\usepackage{jheppub}
\usepackage{amsmath}
\usepackage{graphicx}
\usepackage{multirow}
\usepackage{bm}
\usepackage{bbm}
\usepackage{color}
\usepackage{slashed}
\usepackage{scalefnt}
\usepackage{grffile}

\allowdisplaybreaks

\title{Three-loop matching of  heavy flavor-changing (axial-)tensor  currents}
\author[a]{Wei Tao}
\author[a]{and Zhen-Jun Xiao\footnote{Corresponding author} }
\affiliation[a]{Department of Physics and Institute of Theoretical Physics, Nanjing Normal University, Nanjing 210023, China}
\emailAdd{taowei@njnu.edu.cn}
\emailAdd{xiaozhenjun@njnu.edu.cn}

\abstract{
We present the  three-loop calculations of the  nonrelativistic QCD (NRQCD) current renormalization constants and corresponding anomalous dimensions, and the matching coefficients for the spatial-temporal tensor and spatial-spatial axial-tensor currents with two different heavy quark masses.
We obtain the convergent decay constant ratio up to the next-to-next-to-next-to-leading order (N$^3$LO) for the $S$-wave vector meson $B_c^*$  involving the tensor and axial-tensor currents.
We obtain  the three-loop finite ($\epsilon^0$) term in the ratio of the QCD heavy flavor-changing tensor current renormalization constant  in the on-shell ($\mathrm{OS}$) scheme to that in the   modified-minimal-subtraction   ($\mathrm{\overline{MS}}$) scheme, which is helpful to obtain the three-loop
		 matching coefficients for all  heavy flavor-changing  (axial-)tensor currents.
}

\preprint{~}

\begin{document}

\maketitle

\section{Introduction}

In the Standard Model (SM), the $c\bar b$ meson family is the only meson system  whose states  are formed
from two heavy quarks of different flavors.
As a result, the $c\bar b$ mesons can not annihilate into gluons and consequently they are more stable than the double heavy charmonium $c \bar c$ and  bottomonium $b \bar b$. Therefore, the $c\bar b$ meson family  provides a
good platform for a systematic study of the QCD dynamics in the heavy quark interactions.

The excited $c\bar b$ states can undergo through electromagnetic radiative decays and hadronic transitions   to the  low-lying states, which then decay via the charged weak currents.
Since the $c\bar b$ meson family shares  dynamical properties with the quarkonium, i.e., $c$ and $\bar b$ move nonrelativistically,
it is appropriate to study the low-lying $c\bar b$ meson states by the  NRQCD effective field theory~\cite{Bodwin:1994jh}.
The meson decay constant is a fundamental physical quantity describing   the leptonic decay of a meson state.
With the  framework of  the NRQCD factorization, at the lowest order in quark relative velocity expansion, the decay constant can be factorized into the short-distance coefficient (matching coefficient) and the long-distance matrix element (wave function at the origin).

Using the NRQCD theory, the matching coefficients for  heavy flavor-changing currents have been calculated in various perturbative orders of the QCD strong coupling constant $\alpha_s$.
The  one-loop matching coefficient for the heavy flavor-changing pseudo-scalar current was first calculated in ref.~\cite{Braaten:1995ej}.
The one-loop calculation of the pseudo-scalar and spatial vector currents  allowing for higher order relativistic corrections can be found in refs.~\cite{Hwang:1999fc,Lee:2010ts}.
Two-loop corrections to  the pseudo-scalar and spatial vector currents are available in the literature~\cite{Onishchenko:2003ui,Chen:2015csa,Tao:2022qxa}.
At the N$^3$LO of $\alpha_s$, the matching coefficients for the pseudo-scalar, spatial vector, scalar and spatial axial-vector currents have been numerically evaluated in refs.~\cite{Feng:2022ruy,Sang:2022tnh,Tao:2022hos,Tao:2023mtw}, respectively.

The aim of this work is to calculate the N$^3$LO QCD corrections to the matching coefficients and decay constants for  the $S$-wave vector meson $B_c^*$ coupled with the heavy flavor-changing tensor and axial-tensor currents.
Apart from testing the perturbative convergence of the NRQCD theory, the three-loop matching of  heavy flavor-changing (axial-)tensor  currents will reveal the  nonrelativistic  dynamics  in $B_c^*$ decays involving different currents and  shed light on the internal structure when the heavy bottom and charm quarks are combined into the $c \bar b$ meson.

The  (axial-)tensor   decay constants  can appear in the calculations of meson distribution amplitudes, form factors and branching ratios for the leptonic,   semileptonic, nonleptonic and rare decays~\cite{Hazard:2016fnc,Grinstein:2015aua,Ball:1997rj,Ball:1998kk,Becirevic:2003pn,Yang:2007zt}, which along with experimental measurement are helpful to determine the fundamental parameters in particle physics.
As well as  being significant inputs to  factorization formulae, the (axial-)tensor decay constants  play an important role in QCD sum rule analysis ~\cite{Bell:2010mg,Bakulev:1999gf,Belyaev:1996fd,Broniowski:1998ws}.
Furthermore,  the (axial-)tensor  currents can be included  in effective field theory
extensions of the  SM and may be related to anomalous interactions and new physics beyond the
SM~\cite{Gracey:2022vqr,Blake:2016olu,Chizhov:2003qy}.

By using lattice QCD and QCD sum rules, the (axial-)tensor  decay constants of  heavy quarkonia (such as $\Upsilon$, $J/\psi$) and  light mesons (such as $\rho$, $\phi$ mesons)  have been calculated in various literature ~\cite{Hatton:2020vzp,Braun:2003jg,Bakulev:2000er,Becirevic:1998jp,Becirevic:2013bsa,Ball:1996tb,Ball:2004rg,Ball:2006eu,Craigie:1981jx,Chernyak:1983ej,Govaerts:1986ua,Capitani:1999zd,Jansen:2009yh,Jansen:2009hr,RBC-UKQCD:2008mhs,Cata:2008zc}.
Additionally, the accurate  calculations of  the higher-order perturbative corrections to the  decay constants involving heavy-light (axial-)tensor currents  have been performed  within various QCD    effective field theories~\cite{Broadhurst:1994se,Bell:2010mg,Chetyrkin:2003vi}.
In this paper,  with the help of the NRQCD theory we will fill the gap in the higher-order perturbative QCD calculations for the  (axial-)tensor decay constants of beauty-charmed mesons.
Our predictions for $B_c^*$ decay constants involving vector, tensor and axial-tensor currents will provide valuable information for  experimental searches for the  ground  vector  $B_c^*$ meson.
Additionally,   our calculations will serve as a probe  to test the SM and explore  potential new physics.

The rest of the paper is organized as following.
In Sec.~\ref{Matchingformulas}, we introduce the  matching formulas between QCD and NRQCD.
In Sec.~\ref{QCDvertexfunction},  we describe the details of our calculation for the QCD vertex function.
In Sec.~\ref{ZjQCD}, we study the current renormalization constants in QCD.
In Sec.~\ref{ZjNRQCD}, we  study  the current  renormalization constants   in NRQCD.
In Sec.~\ref{MatchingCandDecayC}, we present the three-loop numeric results of the  matching coefficients and decay constants.
Sec.~\ref{Summary} contains a summary.

\section{Matching formulas ~\label{Matchingformulas}}

We first introduce the definitions of the decay constants for  the $S$-wave vector $c\bar{b}$ meson $B^*_c(1^-)(^3{S_1})$  coupled  with the vector  $v$, tensor $t$,      axial-tensor $t5$ currents~\cite{Bauer:2000yr,Sun:2022hyk,Soni:2017wvy,Colquhoun:2015oha,Dowdall:2012ab,Martin-Gonzalez:2022qwd,Wang:2022cxy,Broadhurst:1994se,Koenigstein:2016tjw,Wang:2012kw,Burakovsky:1997ci,Sundu:2011vz,Abreu:2020ttf,Lu:2006fr,Becirevic:2013bsa,Dhargyal:2016kwp,Aliev:1992vp}
\begin{align}\label{decayCdef}
\langle 0 |j_v^{\mu} |B^*_c(q,\varepsilon) \rangle &\doteq f^{v,i}_{B_c^*}m_{B_c^*} \varepsilon^\mu,   \nonumber\\
\langle 0 |j_t^{\mu\nu} |B^*_c(q,\varepsilon) \rangle &\doteq f^{t,i0} _{B_c^*}(q^\mu \varepsilon^\nu-q^\nu \varepsilon^\mu),   \nonumber\\
\langle 0 |j_{t5}^{\mu\nu} |B^*_c(q,\varepsilon) \rangle &\doteq f^{t5,ij}_{B_c^*}
\epsilon^{\mu\nu\alpha\beta}q_\alpha \varepsilon_\beta,
\end{align}
where $q$ and $\varepsilon$ represent the momentum and  polarization vector of $B_c^*$, respectively.   The superscript $(v,i)/(t,i0)/(t5,ij)$ denotes the contributing (see below) spatial/spatial-temporal/ spatial-spatial component of the vector/tensor/axial-tensor current
, respectively.
The heavy flavor-changing  currents in the full QCD are defined by
\begin{align}\label{QCDcurrents}
j_v^\mu =& \bar{\psi}_b \gamma^\mu \psi_c,
\nonumber \\
j_t^{\mu\nu} =& \bar{\psi}_b \sigma^{\mu\nu}\psi_c,
\nonumber \\
j_{t5}^{\mu\nu} =& \bar{\psi}_b  \sigma^{\mu\nu}\gamma_5 \psi_c,
\end{align}
where $\sigma_{\mu\nu}=\frac{\rm i}{2}(\gamma_{\mu}\gamma_{\nu}-\gamma_{\nu}\gamma_{\mu})$.
The QCD current components contributing to the decay constants of $B_c^*$   can be expanded  in terms of NRQCD currents as follows,
\begin{align}\label{expandcurrents}
j_v^i = &\mathcal{C}_{v,i}\tilde{j}_v^i + {\mathcal O}(|\vec{k}|^{2}),
\nonumber \\
j_t^{i0}    =& \mathcal{C}_{t,i0}\tilde{j}_t^{i0} + {\mathcal O}(|\vec{k}|^{2}),
\nonumber \\
j_{t5}^{ij}    =& \mathcal{C}_{t5,ij}\tilde{j}_{t5}^{ij} + {\mathcal O}(|\vec{k}|^{2}),
\end{align}
where  $|\vec{k}|$  is the small half relative spatial  momentum  between the bottom and charm quarks. $\mathcal{C}_{v,i},\mathcal{C}_{t,i0},\mathcal{C}_{t5,ij}$ are the matching coefficients for the heavy flavor-changing spatial vector, spatial-temporal tensor, spatial-spatial axial-tensor currents, respectively.
And the NRQCD currents read~\cite{Hwang:1999fc,Piclum:2007an}
\begin{align}\label{NRQCDcurrents}
\tilde{j}_v^i = &\varphi_b^\dagger \sigma^i \chi_c,
\nonumber \\
\tilde{j}_{t}^{i0} =&\varphi_b^\dagger{\rm i}   \sigma^i \chi_c,
\nonumber \\
\tilde{j}_{t5}^{ij} =&\varphi_b^\dagger{\rm i}   \sigma^i \sigma^j\chi_c,
\end{align}
where  $\varphi_b^\dagger$ and $\chi_c$ denote 2-component Pauli spinor  fields annihilating the $\bar{b}$ and $c$ quarks, respectively,

After inserting the currents in eq.~\eqref{expandcurrents} between the vacuum state  and the free $c\bar{b}$ pair of on-shell heavy charm and bottom quarks with small relative velocity~\cite{Beneke:1997jm,Onishchenko:2003ui}, we can write the matching formulas as
\begin{align}
\sqrt{Z_{2,b}^\mathrm{OS} Z_{2,c}^\mathrm{OS} } \,Z_{J}^{\mathrm{OS}}  \, \Gamma_{J} =&
\mathcal{C}_{J}(\mu_f,\mu,m_b,m_c) \, \sqrt{\widetilde{Z}_{2,b}^\mathrm{OS} \widetilde{Z}_{2,c}^\mathrm{OS} } \,
{\widetilde Z}_{J}^{-1} \, \widetilde{\Gamma}_{J}  + {\mathcal O}(|\vec{k}|^2),\label{matchingOS} \\
\sqrt{Z_{2,b}^\mathrm{OS} Z_{2,c}^\mathrm{OS} } \,Z_{J}^{\overline{\mathrm{MS}}}  \, \Gamma_{J} =&
\overline{\mathcal{C}}_{J}(\mu_f,\mu,m_b,m_c) \, \sqrt{\widetilde{Z}_{2,b}^\mathrm{OS} \widetilde{Z}_{2,c}^\mathrm{OS} } \,
{\widetilde Z}_{J}^{-1} \, \widetilde{\Gamma}_{J}  + {\mathcal O}(|\vec{k}|^2),\label{matchingMS}
\end{align}
where the contributions from soft, potential and ultrasoft regions of loop momenta have dropped out of both QCD and NRQCD so that $\Gamma_{J}$ is the on-shell unrenormalized  vertex function in  the hard region of QCD~\cite{Beneke:1997jm,Onishchenko:2003ui} while  $\widetilde{\Gamma}_{J}$ is the on-shell tree level vertex function independent of $\alpha_s$ in NRQCD due to the absence of loop contributions on the effective theory.
The left and right parts in the equations represent the renormalization of ${\Gamma}_{J}$ and  $\widetilde{\Gamma}_{J}$, respectively.

$Z_{2,b(c)}^\mathrm{OS}$ is $b(c)$ quark field  $\mathrm{OS}$ renormalization constant in QCD, which can be obtained from refs.~\cite{Fael:2020bgs,Duhr:2019tlz}.
${\widetilde Z}_{2,b(c)}^\mathrm{OS}$ is $b(c)$ quark field  $\mathrm{OS}$ renormalization constant in NRQCD and $\widetilde{Z}_{2,b}^\mathrm{OS}=\widetilde{Z}_{2,c}^\mathrm{OS}=1$ because  heavy bottom and charm quarks are decoupled in the NRQCD effective theory.
${\widetilde Z}_{J}$ is NRQCD  heavy flavor-changing  current  renormalization constant in the   ${\overline{\mathrm{MS}}}$ scheme.
$Z_{J}^{\mathrm{OS}({\overline{\mathrm{MS}}})}$ is the QCD heavy flavor-changing current  renormalization constant in $\mathrm{OS}({\overline{\mathrm{MS}}}$) scheme.

At the leading-order (LO) of $\alpha_s$, the matching coefficient  $\mathcal{C}_{J}^{\text{LO}}=\overline{\mathcal{C}}_{J}^{\text{LO}}=1$,
while in a fixed high order perturbative calculation,
both $\mathcal{C}_{J}$ and $\overline{\mathcal{C}}_{J}$ are finite and depend on the NRQCD factorization scale $\mu_f$ and the QCD  renormalization scale $\mu$.
For $J\in\{(t,i0),(t5,ij)\}$, we can not directly calculate $\mathcal{C}_{J}$ by eq.~\eqref{matchingOS} because both $Z_{J}^{\mathrm{OS}}$ and ${\widetilde Z}_{J}$ are not known at present, however we can obtain $\mathcal{C}_{J}$ by first introducing eq.~\eqref{matchingMS} and calculating $\overline{\mathcal{C}}_{J}$, which will be elucidated in Sec.~\ref{ZjQCD}.

\section{QCD vertex function~\label{QCDvertexfunction}}

Let $q_1(q_2)$ denote the charm (bottom)  external momentum,  $q=q_1+q_2$ represent the total external momentum, and the small momentum $k$~\cite{Zhu:2017lqu} refer to the relative movement  between the bottom and charm quarks.
From eq.~\eqref{expandcurrents} and eq.~\eqref{NRQCDcurrents}, terms at $\mathcal{O}{(k)}$ are not needed in QCD and NRQCD so that we can safely set $k=0$  throughout the calculation to obtain the  vertex function $\Gamma_J$  in the hard region of the full QCD~\cite{Beneke:1997jm}.
Based on the on-shell condition $q_1^2=m_c^2,q_2^2=m_b^2$, the   external momentum configuration can be written as
\begin{align}
q_1=~&\frac{m_c}{m_b+m_c}q,
\nonumber\\
q_2=~&\frac{m_b}{m_b+m_c}q,
\nonumber\\
q^2=~&(m_b+m_c)^2.
\end{align}

Following the literature~\cite{Kniehl:2006qw}, we employ the appropriate projector to obtain the hard QCD vertex function $\Gamma_J$
\begin{align}
\Gamma_{t,i0} = &\mbox{Tr}\left[ {P_{(t,i0),\mu\nu} \Gamma_{(t)}^{\mu\nu}} \right]\,,
\nonumber\\
\Gamma_{t5,ij} = &\mbox{Tr}\left[ {P_{(t5,ij),\mu\nu} \Gamma_{(t5)}^{\mu\nu}} \right]\,,
\end{align}
where  $\Gamma_{(t)}^{\mu\nu}=\cdots\sigma^{\mu\nu}\cdots$, $\Gamma_{(t5)}^{\mu\nu}=\cdots\sigma^{\mu\nu}\gamma_5\cdots$    denote on-shell amputated QCD amplitudes with tensor structures for the tensor and  axial-tensor currents, respectively.
And the projectors for the heavy flavor-changing  spatial-temporal tensor and spatial-spatial axial-tensor currents are constructed as
\begin{align}
P_{(t,i0),\mu\nu} =& \frac{1}{4(D-1)(m_b+m_c)^2}
\left(\frac{m_c}{m_b+m_c}\slashed{q} + m_c
\right) \sigma_{\mu\nu}\left(-\frac{m_b}{m_b+m_c}\slashed{q} + m_b
\right),
\nonumber\\
P_{(t5,ij),\mu\nu} =&
\frac{1}{2(D-1)(D-2)(m_b+m_c)^2}
\left(\frac{m_c}{m_b+m_c}\slashed{q} + m_c
\right) \sigma_{\mu\nu}\gamma_5 \left(-\frac{m_b}{m_b+m_c}\slashed{q} + m_b
\right).
\end{align}
It is worth  mentioning that due to  no singlet diagram~\cite{Piclum:2007an}  and no trace with an odd number of $\gamma_5$~\cite{Tao:2023mtw} for heavy  flavor-changing currents, throughout our calculation we  adopt the naively anticommuting $\gamma_5$ dimensional regularization scheme, i.e., $\gamma_5\gamma_\mu+\gamma_\mu\gamma_5=0,\gamma_5^2={\bf 1}$.

\begin{figure}[htbp]
	\center{
		\includegraphics*[scale=0.9]{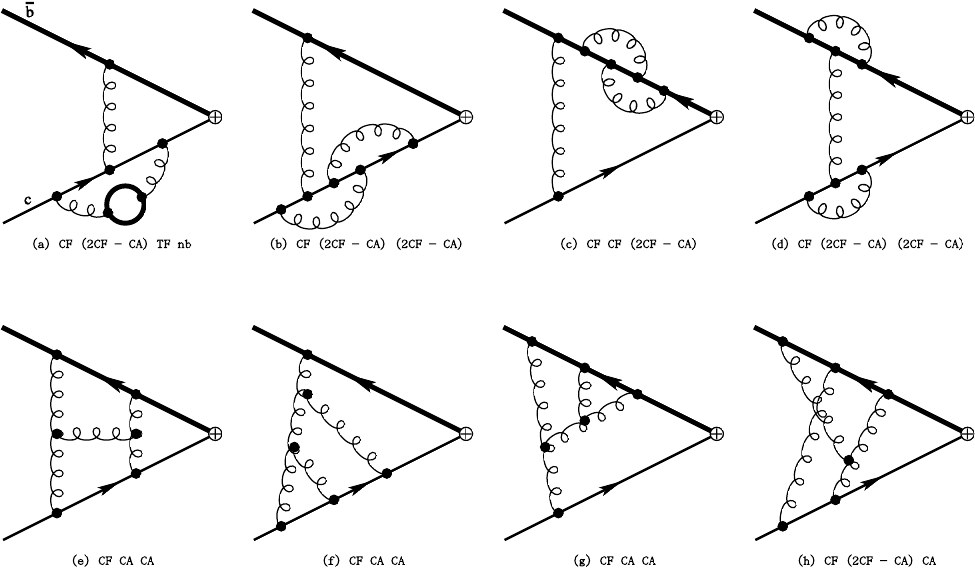}
		\caption {\label{fig:pic3loop} Representative three-loop Feynman diagrams  labelled with corresponding color factors 	for the QCD vertex function with the heavy flavor-changing  current.	 The cross ``$\bigoplus$'' implies the insertion of a certain heavy flavor-changing   current.
			The  solid	closed circle represents the bottom quark loop with mass $m_b$ and flavors $n_b$ (physically, $n_b=1$).}}
\end{figure}

As following, we will outline our workflow to perform the higher-order calculation for the  QCD vertex function.
Firstly, we  use {\texttt{FeynCalc}~\cite{Shtabovenko:2020gxv}}  to obtain Feynman diagrams and corresponding Feynman amplitudes.
In the Feynman diagrams, we have allowed for $n_b$ bottom quarks with mass $m_b$, $n_c$ charm quarks  with mass $m_c$ and $n_l$ massless quarks appearing in the quark loop.
Some representative three-loop Feynman diagrams contributing to the  QCD vertex function  are displayed in figure~\ref{fig:pic3loop}.
By {\texttt{\$Apart}~\cite{Feng:2012iq}}, each Feyman amplitude is decomposed into several Feynman integral families.
Based on the symmetry among different families, we use   {our \texttt{Mathematica} code}+{\texttt{LiteRed}~\cite{Lee:2013mka}}+{\texttt{FIRE6}~\cite{Smirnov:2019qkx}}  to  minimize~\cite{Fael:2020njb,Shtabovenko:2021hjx,Gerlach:2022qnc}  the number of all Feynman integral  families.
For each heavy flavor-changing current, the total number of three-loop Feynman integral families is minimized from 841 to 110. 
Then, we use {\texttt{FIRE6}}/{\texttt{Kira}~\cite{Klappert:2020nbg}}/{\texttt{FiniteFlow}~\cite{Peraro:2019svx}} based on Integration by Parts (IBP)~\cite{Chetyrkin:1981qh} to reduce each Feynman integral family to master integral family.
Next, we use {our \texttt{Mathematica} code}+{\texttt{Kira}}+{\texttt{FIRE6}}  to  minimize the number of  all   master integral families.
For each heavy flavor-changing current, the total number of three-loop master integral families is minimized from 110 to 26 meanwhile the total number of three-loop master integrals is minimized into 300.
Last, we use {\texttt{AMFlow}~\cite{Liu:2022chg}}, which is a proof-of-concept implementation of the auxiliary mass flow method~\cite{Liu:2017jxz}, equipped with {\texttt{FiniteFlow}/\texttt{Kira}} to calculate each master integral family.

\section{QCD current renormalization constants~\label{ZjQCD}}

Based on the matching formulas in eq.~\eqref{matchingOS} and eq.~\eqref{matchingMS}, we have the following relations
for the QCD heavy flavor-changing spatial-temporal  tensor $(t,i0)$ and spatial-spatial axial-tensor  $(t5,ij)$  current
 ${\mathrm{OS}(\mathrm{\overline{MS}})}$ renormalization constants:
\begin{align} \label{zt}
&Z_{t,i0}^{\mathrm{\overline{MS}}}=Z_{t5,ij}^{\mathrm{\overline{MS}}}=Z_t^{\mathrm{\overline{MS}}},\nonumber\\
&Z_{t,i0}^{\mathrm{OS}}=Z_{t5,ij}^{\mathrm{OS}}=Z_t^\mathrm{OS},\nonumber \\
&\frac{\mathcal{C}_{t,i0}}{\overline{\mathcal{C}}_{t,i0}}=\frac{\mathcal{C}_{t5,ij}}{\overline{\mathcal{C}}_{t5,ij}}=\frac{Z_{t}^{\mathrm{OS}}}{Z_{t}^{\mathrm{\overline{MS}}}}=z_t^g z_t^\mu+{\mathcal O}(\epsilon),
\end{align}
where  $Z_t^{\mathrm{OS}(\mathrm{\overline{MS}})}$ is the QCD heavy flavor-changing tensor current ${\mathrm{OS}(\mathrm{\overline{MS}})}$ renormalization constant and $z_t^gz_t^\mu$ is the finite ($\epsilon^0$) term of the ratio $Z_t^{\mathrm{OS}}/Z_t^{\mathrm{\overline{MS}}}$. $Z_t^{\mathrm{OS}}$   is not  available in the literature
 while $Z_t^{\mathrm{\overline{MS}}}$ can be obtained from refs.~\cite{Gracey:2000am,Broadhurst:1994se,Bell:2010mg,Baikov:2006ai,Gracey:2022vqr}:

\begin{align}\label{ZtQCDMS}
Z_t^{\mathrm{\overline{MS}}}=&
1+\frac{\alpha_{s}^{(n_f)}(\mu)}{\pi}\frac{C_F}{4\epsilon}
+\left(\frac{\alpha_{s}^{(n_f)}(\mu)}{\pi}\right)^2 C_F\bigg[C_F\left(\frac{1}{32\epsilon^2}-\frac{19}{64\epsilon}\right)
\nonumber\\&
+C_A\left(-\frac{11}{96\epsilon^2}+\frac{257}{576\epsilon}\right)+T_F n_f\left(\frac{1}{24\epsilon^2}-\frac{13}{144\epsilon}\right)\bigg]
\nonumber\\&
+\left(\frac{\alpha_{s}^{(n_f)}(\mu)}{\pi}\right)^3 C_F\bigg\{C_F^2\bigg[\frac{1}{384\epsilon^3}-\frac{19}{256\epsilon^2}+\frac{1}{\epsilon}\left(\frac{365}{1152}-\frac{1}{3}\zeta_3\right)\bigg]
\nonumber\\&
+C_FC_A\bigg[-\frac{11}{384\epsilon^3}+\frac{75}{256\epsilon^2}+\frac{1}{\epsilon}\left(-\frac{6823}{6912}+\frac{7}{12}\zeta_3\right)\bigg]
\nonumber\\&
+C_A^2\bigg[\frac{121}{1728\epsilon^3}-\frac{3439}{10368\epsilon^2}+\frac{1}{\epsilon}\left(\frac{13639}{20736}-\frac{5}{24}\zeta_3\right)\bigg]
\nonumber\\&
+C_F T_F n_f\bigg[\frac{1}{96\epsilon^3}-\frac{13}{192\epsilon^2}+\frac{1}{\epsilon}\left(\frac{49}{864}+\frac{1}{12}\zeta_3\right)\bigg]
\nonumber\\&
-C_A T_F n_f\bigg[\frac{11}{216\epsilon^3}-\frac{245}{1296\epsilon^2}+\frac{1}{\epsilon}\left(\frac{251}{1296}+\frac{1}{12}\zeta_3\right)\bigg]
\nonumber\\&
+T_F^2 n_f^2\bigg[\frac{1}{108\epsilon^3}-\frac{13}{648\epsilon^2}-\frac{1}{144\epsilon}\bigg]
\bigg\}+\mathcal{O}\left(\alpha_s^4\right).
\end{align}

On the one hand, we can use eq.~\eqref{matchingMS} and eq.~\eqref{ZtQCDMS} to fit ${\widetilde Z}_J$ and calculate $\overline{\mathcal{C}}_J$ for $J\in\{(t,i0),(t5,ij)\}$.
On the other hand, from eq.~\eqref{expandcurrents} and eq.~\eqref{NRQCDcurrents}, we obtain  following relations between the spatial vector and spatial-temporal tensor currents:
\begin{align}\label{currentrelations}
{\widetilde Z}_{t,i0}=&{\widetilde Z}_{v,i},\nonumber\\
\mathcal{C}_{t,i0}=&\mathcal{C}_{v,i},\nonumber\\
f_{B_c^*}^{t,i0}=&f_{B_c^*}^{v,i},
\end{align}
where ${\widetilde Z}_{v,i}$, $\mathcal{C}_{v,i}$ and $f_{B_c^*}^{v,i}$
have been calculated and denoted as ${\widetilde Z}_{v}$, $\mathcal{C}_{v}$ and $f_{B_c^*}$
 respectively in our previous publication~\cite{Tao:2023mtw}.
Substituting eq.~\eqref{currentrelations} into eq.~\eqref{zt}, we obtain
\begin{align}
z_t^g z_t^\mu=\frac{\mathcal{C}_{v,i}}{\overline{\mathcal{C}}_{t,i0}}.
\end{align}
For $J\in\{(t,i0),(t5,ij)\}$,
with $z_t^gz_t^\mu$ and ${\overline{\mathcal{C}}_{J}}$ known, we can calculate ${\mathcal{C}_{J}}$ by eq.~\eqref{zt}, i.e. ${\mathcal{C}_{J}}=z_t^gz_t^\mu{\overline{\mathcal{C}}_{J}}$.

As following, we will present our  result of $z_t^gz_t^\mu$. For brevity, we introduce several notations throughout the paper:
\begin{align}
x\equiv& {m_c\over m_b},
\nonumber\\
L_{\mu}\equiv& \ln \frac{\mu^2}{m_b m_c},
\nonumber\\
L_{\mu_f}\equiv& \ln \frac{\mu_f^2}{m_b m_c}.
\end{align}

Let~\footnote{We find that ${\mathcal{C}_{J}}/{z_t^g}={z_t^\mu} {\overline{\mathcal{C}}_{J}}\,(J\in\{(t,i0),(t5,ij)\})$ is renormalization group invariant and $z_t^gz_t^\mu$ can be written as $z_t^gz_t^\mu=\sum_{0\leq j\leq i} \left(\alpha_s^{(n_f)}(\mu)/\pi\right)^i L_\mu^j c_{ij}(x)$, which can always be factorized into the product of $z_t^\mu=1+\sum_{1\leq j\leq i}\left(\alpha_s^{(n_f)}(\mu)/\pi\right)^i L_\mu^j f_{ij}(x)$ and the renormalization group invariant $z_t^g$ in eq.~\eqref{ztg}. In a word, $z_t^\mu$ and $z_t^g$ can be uniquely determined by    ${\overline{\mathcal{C}}_{t,i0}}$ and ${\mathcal{C}_{t,i0}}={\mathcal{C}_{v,i}}$.} ${z_t^{\mu}}({L_\mu=0})=1$, and let $z_t^g$ satisfy the renormalization group invariance (see eq.~(5.14) in ref.~\cite{Tao:2023mtw}).
With the aid of numerical fitting techniques such as the PSLQ algorithm~\cite{Duhr:2019tlz}, we can obtain the following expressions for $z_t^\mu$ and $z_t^g$:
\begin{align}
\label{ztmu}
z_t^\mu=&1+\frac{\alpha_{s}^{(n_f)}(\mu)}{\pi}\frac{C_F}{4}  L_{\mu }+\left(\frac{\alpha_{s}^{(n_f)}(\mu)}{\pi}\right)^2
C_F \bigg[ C_F \left(\frac{1}{32}L_{\mu }^2-\frac{19 }{32}L_{\mu }\right)
\nonumber\\&
+C_A \left(\frac{11 }{96}L_{\mu }^2+\frac{257 }{288}	L_{\mu }\right) - T_F n_f\left(\frac{1}{24}L_{\mu }^2+\frac{13 }{72}L_{\mu }\right)\bigg]
\nonumber\\&
+\left(\frac{\alpha_{s}^{(n_f)}(\mu)}{\pi}\right)^3 C_F \bigg\{ C_F^2 \bigg[\frac{1}{384}L_{\mu }^3-\frac{19 }{128}L_{\mu
}^2+\left(\frac{365}{384}-\zeta_3\right) L_{\mu }\bigg]\nonumber\\&
+ C_F C_A \bigg[\frac{11}{384} L_{\mu }^3-\frac{185}{576} L_{\mu}^2+\left(\frac{7}{4} \zeta_3-\frac{6823}{2304}\right) L_{\mu}\bigg]
\nonumber\\&
+C_A^2 \bigg[\frac{121}{1728} L_{\mu }^3+\frac{3133}{3456} L_{\mu}^2+\left(\frac{13639}{6912}-\frac{5}{8} \zeta_3\right) L_{\mu}\bigg]
\nonumber\\&
+C_F T_F n_f \bigg[-\frac{1}{96}L_{\mu }^3+\frac{35}{288} L_{\mu}^2+\left(\frac{\zeta_3}{4}+\frac{49}{288}\right) L_{\mu}\bigg]
\nonumber\\&
-C_A  T_F n_f \bigg[\frac{11}{216} L_{\mu }^3+\frac{445}{864} L_{\mu}^2+\left(\frac{\zeta_3}{4}+\frac{251}{432}\right) L_{\mu }\bigg]
\nonumber\\&
+T_F^2 n_f^2\bigg[\frac{L_{\mu }^3}{108}+\frac{13 L_{\mu }^2}{216}-\frac{L_{\mu }}{48}\bigg] \bigg\}+\mathcal{O}\left(\alpha_s^4\right),
\\
\label{ztg}
z_t^g=&1+\frac{\alpha_s^{(n_f)}(\mu)}{\pi} z_t^{(1)}(x) +\left(\frac{\alpha_s^{(n_f)}(\mu)}{\pi}\right)^2
\left(z_t^{(2)}(x)+\frac{z_t^{(1)}(x)}{4}\beta_0^{(n_f)}L_{\mu } \right)
\nonumber \\&
+\left(\frac{\alpha_s^{(n_f)}(\mu)}{\pi}\right)^3\Bigg\{ z_t^{(3)}(x)
+\left(\frac{z_t^{(1)}(x)}{16}\beta_1^{(n_f)}+\frac{z_t^{(2)}(x)}{2}\beta_0^{(n_f)}\right)L_{\mu }
\nonumber \\&
+ \frac{z_t^{(1)}(x)}{16}{\beta_0^{(n_f)}}^2L_{\mu }^2  \Bigg\} +\mathcal{O}\left(\alpha_s^4\right),
\\
z_t^{(1)}(x) =&-\frac{C_F}{4} \, \frac{x-1 }{x+1}\ln x,
\nonumber\\
z_t^{(2)}(x) =& C_F \,\Big[ C_F \,z_t^{FF}(x)+ C_A\, z_t^{FA}(x)
+ T_F\, n_l\, z_t^{FL}(x)+ T_F\, n_b \,z_t^{FB}(x) + T_F\, n_c\, z_t^{FC}(x)\Big],
\nonumber\\
z_t^{(3)}(x) =& C_F\,\Big[ C^2_F \, z_t^{FFF}(x)+C_F \,C_A \, z_t^{FFA}(x)+ C_A^2 \, z_t^{FAA}(x)
\nonumber\\&
+C_F\, T_F\, n_l\, z_t^{FFL}(x)+ C_F\,  T_F\, n_b\, z_t^{FFB}(x)+C_F\,T_F \, n_c\,  z_t^{FFC}(x)
\nonumber\\&
+C_A\,T_F\, n_l\,z_t^{FAL}(x) + C_A\,T_F\, n_b\, z_t^{FAB}(x)  + C_A\,T_F \, n_c\, z_t^{FAC}(x)
\nonumber\\&
+T_F^2\, n_l^2\, z_t^{FLL}(x)+T_F^2 \,n_l\, n_b \,   z_t^{FLB}(x)+T_F^2 \,n_l\,  n_c \,  z_t^{FLC}(x)
\nonumber\\&
+T_F^2\, n_b^2\, z_t^{FBB}(x)+ T_F^2 \, n_b \, n_c \, z_t^{FBC}(x) +T_F^2 \, n_c^2\,  z_t^{FCC}(x)
\Big],
\end{align}
where $\beta_0^{(n_f)}=\frac{11}{3}C_A-\frac{4}{3} T_F n_f$ and  $\beta_1^{(n_f)}=\frac{34}{3}C_A^2-4 C_F T_F n_f-\frac{20}{3} C_A T_F n_f$ are respectively the one-loop and two-loop  coefficients of the QCD $\beta$ function~\cite{vanRitbergen:1997va} and $n_f=n_l+n_b+n_c$ is the total number of flavors.
The color-structure components of $z_t^{(2)}(x)$ and $z_t^{(3)}(x)$ read:
\begin{align}
z_t^{FF}(x)=&-\frac{563}{384}-\frac{1}{6} \pi ^2\ln 2+\frac{\zeta_3 }{4}
+\frac{3 (x-1) }{32 (x+1)}\ln x
\nonumber\\&
-\frac{ 8 x^4-20 x^3-99 x^2-46 x-35}{144
	(x+1)^2}\pi ^2
\nonumber\\&
-\frac{32 x^4+40 x^3-19 x^2+42 x-3}{96 (x+1)^2}\ln^2 x
\nonumber\\&
+ \frac{(x+1) (x-1)^3}{3 x^2} \big[\ln (1-x) \ln x+\text{Li}_2(x)\big]
\nonumber\\&
+\frac{2 x^4+x^3-x-2}{6	x^2}\big[\ln (1+x)\ln x +\text{Li}_2(-x)\big],
\nonumber\\
z_t^{FA}(x)=&  \frac{5141}{3456}+ \frac{1}{12} \pi ^2 \ln 2 -\frac{\zeta_3}{8} -\frac{209 (x-1) }{288	(x+1)}\ln x
\nonumber\\&
+\frac{ x^4-2 x^3-10 x^2-4 x-3 }{36 (x+1)^2}\pi ^2
\nonumber\\&
+\frac{16 x^4+16 x^3-5 x^2+22 x+11}{96 (x+1)^2} \ln
^2 x
\nonumber\\&
-\frac{(x+1) (x-1)^3}{6 x^2} \big[\ln (1-x) \ln x+\text{Li}_2 (x) \big]
\nonumber\\&
-\frac{ x^4-1 }{6 x^2}\big[ \ln (1+x)\ln x+\text{Li}_2(-x)\big],
\nonumber\\
z_t^{FL}(x)=& -\frac{205}{864}-\frac{\pi ^2}{36}+\frac{13 (x-1) }{72
	(x+1)}\ln x-\frac{1}{24} \ln^2 x,
\nonumber\\
z_t^{FB}(x)=&-\frac{205}{864}-\frac{1}{4 x}+\frac{\pi ^2}{18	(x+1)}
\nonumber\\&
+\frac{ 13 x^2-13 x+12 }{72 x (x+1)}\ln x+\frac{(3 x-1) }{24 (x+1)}\ln ^2 x
\nonumber\\&
-\frac{ (x^2+x+1 )  (x-1)^2}{6 x^3 (x+1)}\big[\ln (1-x) \ln x
+\text{Li}_2(x)\big]
\nonumber\\&
 -\frac{ x^3+1 }{6 x^3}\big[ \ln (x+1)\ln x+\text{Li}_2(-x)\big],
\nonumber \\
z_t^{FC}(x)=&-\frac{205}{864}-\frac{x}{4} -\frac{  x^4-3 x^3-5 x+1 }{36	(x+1)}\pi ^2
\nonumber\\&
-\frac{12 x^2-13 x+13}{72 (x+1)} \ln x-\frac{ 4 x^4+x+1 }{24 (x+1)}\ln ^2 x
\nonumber\\&
+ \frac{(x^2+x+1)  (x-1)^2}{6 (x+1)}\big[\ln (1-x) \ln x+\text{Li}_2(x)\big]
\nonumber\\&
+\frac{x^3+1}{6} \big[ \ln(x+1)\ln x +\text{Li}_2(-x)\big],
\\
z_t^{FFF}(x_0)=&-2.322282618854114578537016108614,
\nonumber\\
z_t^{FFA}(x_0)=&0.63952094914985889385999778652907,
\nonumber\\
z_t^{FAA}(x_0)=& 4.91543462857763455194218954249917,
\nonumber\\
z_t^{FFL}(x_0)=& 0.48535345668429975701679412257185,
\nonumber\\
z_t^{FFB}(x_0)=&-0.96788752784853089190831478824595,
\nonumber\\
z_t^{FFC}(x_0)=&-0.030004714341714672058240640021062,
\nonumber\\
z_t^{FAL}(x_0)=& -3.7810411909098785095485086146338,
\nonumber\\
z_t^{FAB}(x_0)=& 2.00151369570156466157964680499125,
\nonumber\\
z_t^{FAC}(x_0)=&-0.64029413850834349156677913068544,
\nonumber\\
z_t^{FLL}(x)=&\frac{2665}{23328} +\frac{13 \pi ^2}{324}+\frac{7 \zeta_3}{54}-\frac{89 (x-1) }{648 (x+1)}\ln x
\nonumber\\&
-\frac{ x-1 }{54 (x+1)}\pi^2\ln x +\frac{13}{216} \ln^2 x -\frac{x-1 }{108 (x+1)} \ln^3 x,
\nonumber\\
z_t^{FLB}(x_0)=&-0.18426684902451221497413586356109,
\nonumber\\
z_t^{FLC}(x_0)=&0.24724217746243652013783508427839,
\nonumber\\
z_t^{FBB}(x_0)=&0.24641742011807953984404155385694,
\nonumber\\
z_t^{FBC}(x_0)=&0.069455208354306644824678292877365,
\nonumber\\
z_t^{FCC}(x_0)=&0.043546270000908910321578999401716,
\end{align}
where the numerical results with about 30-digit precision for various
color-structure components of $z_t^{(3)}(x)$ at the physical point $x=x_0=150/475$
are presented because it is difficult to obtain the analytic expressions of them involving  Goncharov polylogarithms  (see ref.~\cite{Fael:2020bgs}).
In the ancillary file attached to the paper, we provide the numerical results with about 30-digit precision for  them at the following ten points:
\begin{align}\label{x10}
x\in\bigg\{\frac{1}{20}, \frac{1}{5}, \frac{100}{475}, \frac{150}{525}, \frac{150}{475}, \frac{150}{425}, \frac{204}{498}, \frac{200}{475}, \frac{1}{2}, 1\bigg\}.
\end{align}
The values of them for $x>1$ can be obtained by employing the invariance of $z_t^g$  under the exchange $m_b\leftrightarrow m_c$ meanwhile $n_b\leftrightarrow n_c$.

To verify our calculation of $z_t^g z_t^\mu$ and investigate the deviation between $\mathcal{C}_J$ and $\overline{\mathcal{C}}_J$,
following eq.~\eqref{zt}, we also study the relations~\cite{Chetyrkin:2003vi} for     the QCD heavy flavor-changing scalar $(s)$ and pseudo-scalar $(p)$ current ${\mathrm{OS}(\mathrm{\overline{MS}})}$ renormlization constants:
\begin{align} \label{zm}
&Z_{s}^\mathrm{\overline{MS}}=Z_{p}^\mathrm{\overline{MS}}=Z_m^\mathrm{\overline{MS}},\nonumber \\
&Z_{s}^\mathrm{OS}=Z_{p}^\mathrm{OS}=\frac{m_b Z_{m,b}^\mathrm{OS}+m_c Z_{m,c}^\mathrm{OS}}{m_b+m_c},\nonumber \\
&\frac{\mathcal{C}_{s}}{\overline{\mathcal{C}}_{s}}=\frac{\mathcal{C}_{p}}{\overline{\mathcal{C}}_{p}}=\frac{m_b Z_{m,b}^\mathrm{OS}+m_c Z_{m,c}^\mathrm{OS}}{(m_b+m_c){Z_{m}^{\mathrm{\overline{MS}}}}}
=z_m^g z_m^\mu+{\mathcal O}(\epsilon),
\end{align}
where  $Z_m^\mathrm{\overline{MS}}$ is quark mass $\mathrm{\overline{MS}}$ renormalization constant in QCD, which can be found in refs.~\cite{Marquard:2016dcn,Gracey:2000am,Broadhurst:1994se,Bell:2010mg}.
$Z_{m,b(c)}^\mathrm{OS}$ is $b(c)$ quark mass  $\mathrm{OS}$ renormalization constant in QCD,
which  can be obtained from ref.~\cite{Fael:2020bgs}.
$z_m^g$ and $z_m^\mu$  can be defined by analogizing to the definitions
 of $z_t^g$ and $z_t^\mu$ respectively in the above context.

\begin{figure}[htbp]
	\centering
	\includegraphics[width=0.44\textwidth]{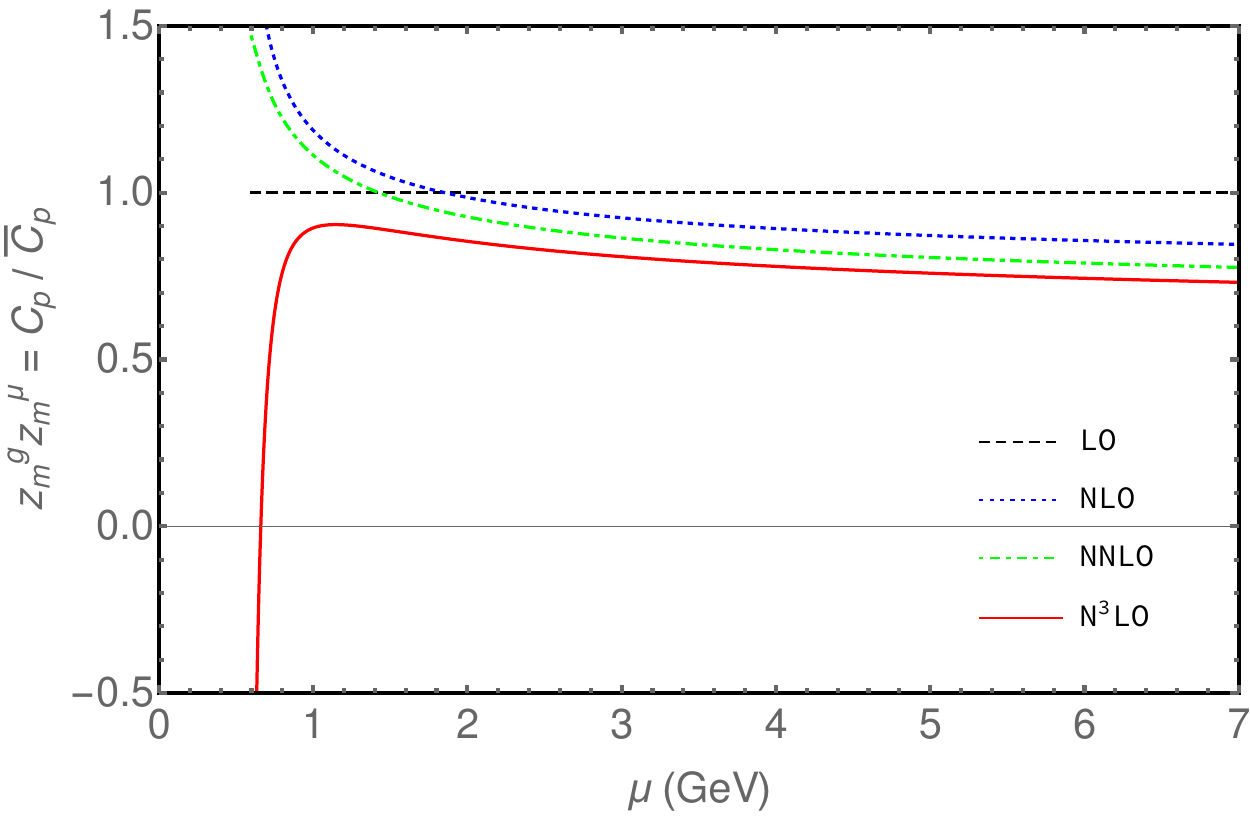}\qquad
	\includegraphics[width=0.44\textwidth]{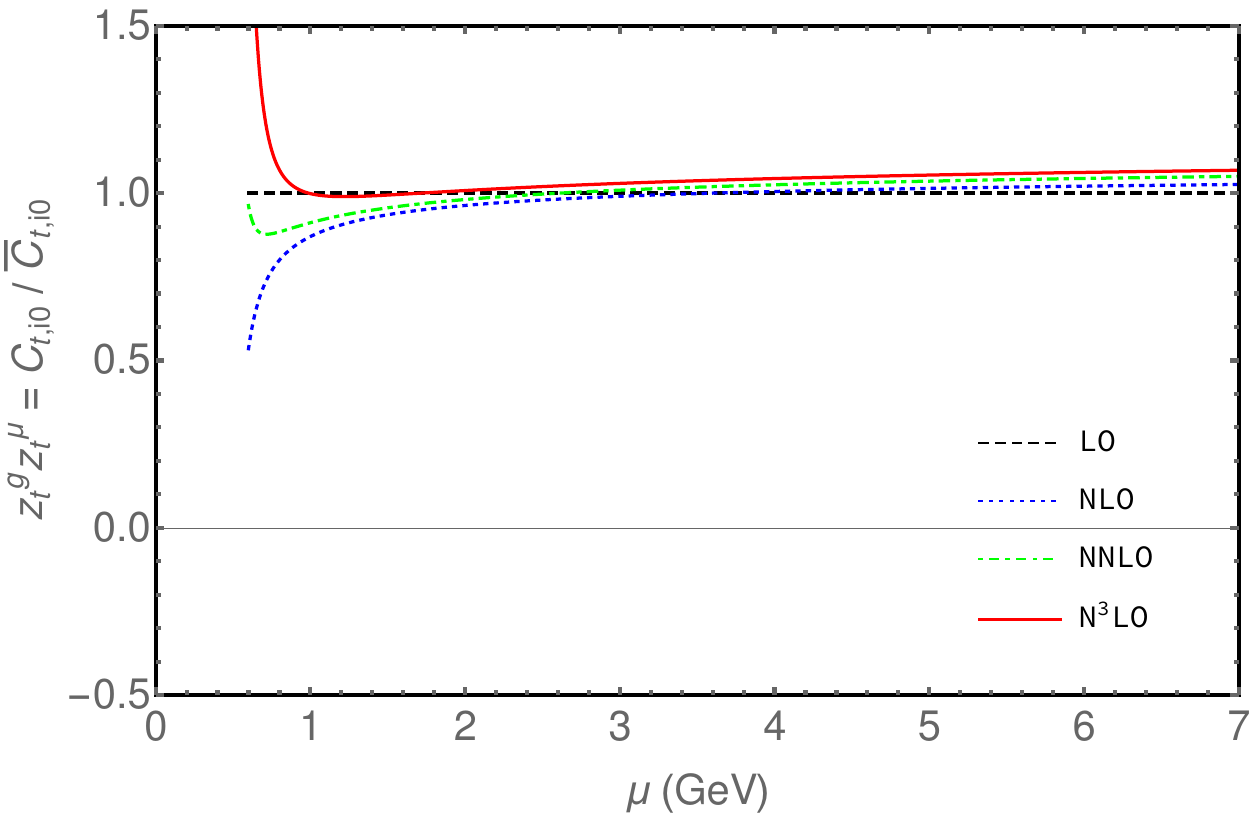}\qquad
	\caption{The renormalization scale $\mu$ dependence of $z_m^g z_m^\mu=\frac{\mathcal{C}_p}{\overline{\mathcal{C}}_p}$ and $z_t^g z_t^\mu=\frac{\mathcal{C}_{t,i0}}{\overline{\mathcal{C}}_{t,i0}}$
		at LO,  NLO,  NNLO and N$^3$LO accuracy. The central values  are calculated inputting the  physical values with  $\mu_f=1.2~\,\mathrm{GeV}$,  $m_b=4.75\mathrm{GeV}$ and $m_c=1.5\mathrm{GeV}$.  There are no visible  error bands  from the variation of the NRQCD factorization scale $\mu_f$  between  7 and 0.4 $\mathrm{GeV}$. }
	\label{fig:zmu}
\end{figure}
Furthermore, we expand $z_m^g z_m^\mu=\frac{\mathcal{C}_p}{\overline{\mathcal{C}}_p}$ and $z_t^g z_t^\mu=\frac{\mathcal{C}_{t,i0}}{\overline{\mathcal{C}}_{t,i0}}$ in power series of $\alpha_s^{(n_l=3)}(\mu)$ (where $n_l$ is the number of massless quark flavors. See the following sections for the definition of $\alpha_s$.) and plot the renormalization scale $\mu$ dependence of them in figure~\ref{fig:zmu}.
We see  both  $z_m^g z_m^\mu$ and $z_t^g z_t^\mu$ are  convergent and show good renormalization  scale   dependence.
Note that both  $z_m^g z_m^\mu$ and $z_t^g z_t^\mu$ are free from  $\mu_f$ due to the fact that the QCD current renormalization constant
$Z_J^{\mathrm{OS}(\mathrm{\overline{MS}})}$ is independent of the NRQCD factorization scale $\mu_f$.
We also find  although $\mathcal{C}_J$ satisfies the renormalization group invariance (see eq.~(5.14) in ref.~\cite{Tao:2023mtw}) while $\overline{\mathcal{C}}_J$ does not, the deviation between $\mathcal{C}_J$ and $\overline{\mathcal{C}}_J$ is relatively small.
In addition, our calculation verifies both $\overline{\mathcal{C}}_{J}$ and $\mathcal{C}_{J}$
are gauge invariant so that  $z_m^g z_m^\mu$, $z_t^g z_t^\mu$, $Z_J^{\mathrm{\overline{MS}}}$ and $Z_J^{\mathrm{OS}}$ are also  gauge invariant.
We conclude that our calculation results for $z_t^g z_t^\mu$ are reasonable and reliable.

\section{NRQCD current renormalization constants~\label{ZjNRQCD}}

We employ the matching formula in eq.~\eqref{matchingMS} to obtain ${\widetilde Z}_J$ for $J\in\{(t,i0),(t5,ij)\}$.
To perform the  conventional QCD renormalization procedure~\cite{Davydychev:1997vh} for {${\Gamma}_J$} on the l.h.s of eq.~\eqref{matchingMS},
we need to implement the QCD heavy quark field and  mass ${\rm OS}$ renormalization,  the QCD coupling constant $\overline{\rm MS}$ renormalization~\cite{Mitov:2006xs,Chetyrkin:1997un,vanRitbergen:1997va},
and the QCD heavy flavor-changing current $\overline{\rm MS}$ renormalization,
after which the QCD vertex function gets rid of the ultra-violet(UV) divergences, yet still contains  uncancelled infra-red(IR) poles starting from order $\alpha_s^2$.
The remaining IR poles in QCD should be exactly  cancelled by the UV poles of the NRQCD heavy flavor-changing current $\overline{\rm MS}$ renormalization constant ${\widetilde Z}_{J}$  on the r.h.s of eq.~\eqref{matchingMS}, which renders the matching coefficient finite. Therefore, eq.~\eqref{matchingMS} can completely determine ${\widetilde Z}_J$ and  subsequently determine $\overline{\mathcal{C}}_J$.

Based on the high-precision numerical results and  the PSLQ algorithm~\cite{Duhr:2019tlz}, we have fitted and reconstructed the exact analytical expressions of ${\widetilde Z}_J$ for $J\in\{(t,i0),(t5,ij)\}$, which  verify ${\widetilde Z}_{t,i0}\equiv{\widetilde Z}_{v,i}$.
The results of ${\widetilde Z}_{t,i0}$  and ${\widetilde Z}_{t5,ij}$  are presented  as following:
\begin{align}
\widetilde{Z}_{J}\left(L_{\mu_f};x\right)
=&1+\left(\frac{\alpha_{s}^{(n_{l})}(\mu_f)}{\pi}\right)^{2}\widetilde{Z}_{J}^{(2)}(x)+\left(\frac{\alpha_{s}^{(n_{l})}(\mu_f)}{\pi}\right)^{3}\widetilde{Z}_{J}^{(3)}\left(L_{\mu_f};x\right)+\mathcal{O}(\alpha_s^4),
\nonumber\\
\widetilde{Z}_{t,i0}^{(2)}(x)=&\widetilde{Z}_{t5,ij}^{(2)}(x)=\pi^{2}C_{F}\frac{1}{\epsilon}\left(\frac{3x^2+2x+3}{24\left(x+1\right)^2}C_{F}+\frac{1}{8}C_{A}\right),
\nonumber\\
\widetilde{Z}_{J}^{(3)}\left(L_{\mu_f};x\right)=&
\pi^{2}C_{F}\bigg\{
C_F^2 \bigg[\frac{3 x^2-x+3}{36\epsilon ^2 (x+1)^2 }
+\frac{1}{\epsilon}\left(
\frac{19 x^2+5 x+19}{36 (x+1)^2}-\frac{2 }{3}\ln 2
\right.\nonumber\\&\left.
+\frac{x^3-4 x^2-2 x-3}{12 (x+1)^3}\ln x
+\frac{1}{6}
\ln (x+1)
+\frac{3 x^2-x+3 }{12 (x+1)^2}L_{\mu _f}
\right)\bigg]
\nonumber\\ &
+
C_F C_A \bigg[ \frac{x}{216\epsilon ^2 (x+1)^2 } + \frac{1}{ \epsilon }\left( \frac{78 x^2+c_1^J\,	x+78}{324 (x+1)^2} \right.
\nonumber\\ &
\left. -\frac{x+11 }{48 (x+1)}\ln x +\frac{1}{4} \ln (x+1) +\frac{11 x^2+8 x+11}{48 (x+1)^2} L_{\mu _f} \right)\bigg]
\nonumber\\ &
+C_A^2  \bigg[\frac{-1}{16 \epsilon ^2}+\frac{1}{ \epsilon}\left(\frac{2}{27}+\frac{1}{6}\ln2-\frac{1}{24}\ln x+\frac{1}{12}\ln (x+1) +
\frac{1}{24}L_{\mu _f}\right)\bigg]
\nonumber\\ &
+ C_F T_F n_l\bigg[\frac{3 x^2+2	x+3}{108\epsilon ^2 (x+1)^2 }-\frac{21 x^2+c_2^J\, x+21}{324 \epsilon (x+1)^2 }\bigg]
+C_F  T_F n_b\frac{ x^2 }{15\epsilon (x+1)^2  }
\nonumber\\ &
+C_F  T_F n_c\frac{ 1 }{15\epsilon (x+1)^2  }+C_A  T_F n_l \bigg[
\frac{1}{36 \epsilon ^2}-\frac{37}{432 \epsilon } \bigg]\bigg\},
\end{align}
where $c_1^{t,i0}=296$, $c_1^{t5,ij}=227$, $c_2^{t,i0}=58$, $c_2^{t5,ij}=10$.
And the corresponding anomalous dimension $\tilde{\gamma}_{J}$~\cite{Groote:1996xb,Kiselev:1998wb,Henn:2016tyf,Fael:2022miw,Grozin:2015kna,Ozcelik:2021zqt}  related to $\widetilde{Z}_{J}$ reads
\begin{align}
\tilde{\gamma}_{J}\left(L_{\mu_f};x\right) =&\left(\frac{\alpha_s^{(n_l)}
	\left(\mu_f\right)}{\pi}\right)^2 \tilde{\gamma}_{J}^{(2)}\left(x\right)+\left(\frac{\alpha_s^{(n_l)}
	\left(\mu_f\right)}{\pi}\right)^3 \tilde{\gamma}_{J}^{(3)}\left(L_{\mu_f};x\right)+\mathcal{O}(\alpha_s^4),
\nonumber\\
\tilde{\gamma}_{J}^{(2)}(x)=&-4\,\widetilde{Z}_{J}^{(2)[1]}(x),
~~~~~~
\tilde{\gamma}_{J}^{(3)}\left(L_{\mu_f};x\right)=-6\,\widetilde{Z}_{J}^{(3)[1]}\left(L_{\mu_f};x\right),
\end{align}
where  $\widetilde{Z}_{J}^{(i)[1]}$ denotes the coefficient of  $\frac{1}{\epsilon}$  in $\widetilde{Z}_{J}^{(i)}$.
Note both $\widetilde{Z}_{J}$ and  $\tilde{\gamma}_{J}$ explicitly depend on  $\mu_f$ but not  $\mu$~\cite{Marquard:2014pea,Egner:2022jot,Feng:2022vvk,Feng:2022ruy,Sang:2022tnh}.
One can check $\widetilde{Z}_{J}$ and $\tilde{\gamma}_{J}$  are invariant   under the exchange $m_b\leftrightarrow m_c$ meanwhile $n_b\leftrightarrow n_c$.

In our calculation, we consider QCD where $n_l$ massless flavors, $n_b$ flavors with mass $m_b$ and $n_c$ flavors with mass $m_c$ possibly appear in the quark loop. However the contributions from the loops of heavy charm  and bottom quarks  are decoupled in the NRQCD.
To match QCD with NRQCD,
we employ both the coupling running~\cite{Abreu:2022cco,Tao:2022hos,Tao:2023mtw} and the decoupling relation~\cite{Chetyrkin:2005ia,Bernreuther:1981sg,Grozin:2011nk,Grozin:2012ic,Barnreuther:2013qvf,Grozin:2007fh,Gerlach:2019kfo,Ozcelik:2021zqt,Tao:2022hos,Tao:2023mtw} in $D=4-2\epsilon$ for the mutual conversion between
 $\alpha_s^{(n_f)}(\mu)$, $\alpha_s^{(n_l)}(\mu_f)$ and $\alpha_s^{(n_l)}(\mu)$, where $n_f=n_l+n_b+n_c$ is the total number of flavors.

The numerical values of $\alpha_s^{(n_l)}(\mu)$  with $n_l=3,\,n_b=n_c=1$ and $\mu\in[0.4,7]\,\mathrm{GeV}$ can be calculated  using  the coupling running and the decoupling relation in $D=4$~\cite{Tao:2022hos,Tao:2023mtw}
or  using the package  {\texttt{RunDec}~\cite{Chetyrkin:2000yt,Schmidt:2012az,Deur:2016tte,Herren:2017osy}} function {\texttt{AlphasLam}}  with  $\Lambda_{QCD}^{(n_l=3)}=0.3344\mathrm{GeV}$
determined    by inputting the initial value  $\alpha_s^{(n_f=5)}\left(m_Z=91.1876\mathrm{GeV}\right)=0.1179$.

\section{Matching coefficients and decay constants~\label{MatchingCandDecayC}}

The final result of the matching coefficient $\mathcal{C}_J$ for $J\in\{(t,i0),(t5,ij)\}$ can be written as~\cite{Feng:2022ruy,Sang:2022tnh,Tao:2022hos,Tao:2023mtw}:
\begin{align}
\label{Cjformula}
&\mathcal{C}_J(\mu_f,\mu,m_b,m_c) = 1+\frac{\alpha_s^{(n_l)}(\mu)}{\pi} \mathcal{C}_J^{(1)}(x)
\nonumber \\&
+\left(\frac{\alpha_s^{(n_l)}(\mu)}{\pi}\right)^2
\bigg[\frac{\mathcal{C}_J^{(1)}(x)}{4}\beta_0^{(n_l)}L_{\mu }
+\frac{\tilde{\gamma}_J^{(2)}(x)}{2}L_{\mu _f}+\mathcal{C}_J^{(2)}(x)\bigg]
\nonumber \\&
 +\left(\frac{\alpha_s^{(n_l)}(\mu)}{\pi}\right)^3\Bigg\lbrace
\frac{\mathcal{C}_J^{(1)}(x)}{16}{\beta_0^{(n_l)}}^2L^2_{\mu}
+\bigg[\frac{\mathcal{C}_J^{(1)}(x)}{16}\beta_1^{(n_l)}+\frac{\mathcal{C}_J^{(2)}(x)}{2}\beta_0^{(n_l)}\bigg]L_{\mu }
\nonumber \\&
+\frac{\tilde{\gamma}_J^{(2)}(x)}{4}\beta_{0}^{(n_l)}L_{\mu }L_{\mu _f}
+\bigg[\frac{\partial\tilde{\gamma}_J^{(3)}\left(L_{\mu _f};x\right)}{4\partial L_{\mu _f}}-\frac{\tilde{\gamma}_J^{(2)}(x)}{8}\beta_0^{(n_l)}\bigg]L^2_{\mu _f}
\nonumber \\&
 +\frac{1}{2}\bigg[\mathcal{C}_J^{(1)}(x) \tilde{\gamma}_J^{(2)}(x)+\tilde{\gamma}_J^{(3)}\left(L_{\mu _f}=0;x\right)\bigg]L_{\mu _f}
+ \mathcal{C}_J^{(3)}(x) \Bigg\rbrace+\mathcal{O}\left(\alpha_s^4\right),
\end{align}
where $n_l$ is the number of the massless flavors.  $\mathcal{C}_{J}^{(n)}(x)~(n=1,2,3)$ is a function only depending on $x=m_c/m_b$, which can be decomposed in terms of different color factor structures~\cite{Marquard:2014pea,Beneke:2014qea,Egner:2022jot,Feng:2022vvk,Feng:2022ruy,Sang:2022tnh}:
\begin{align}
\mathcal{C}_{t,i0}^{(1)}(x)=&\mathcal{C}_{t5,ij}^{(1)}(x)=\frac{3}{4}C_F\left(\frac{x-1}{x+1}\,\ln x-\frac{8}{3}\right),
\nonumber \\
\mathcal{C}_J^{(2)}(x) =& C_F \,\Big[ C_F \,\mathcal{C}_J^{FF}(x)+ C_A\, \mathcal{C}_J^{FA}(x)
+ T_F\, n_l\, \mathcal{C}_J^{FL}(x)+ T_F\, n_b \,\mathcal{C}_J^{FB}(x) + T_F\, n_c\, \mathcal{C}_J^{FC}(x)\Big],
\nonumber\\
\mathcal{C}_J^{(3)}(x) =& C_F\,\Big[ C^2_F \, \mathcal{C}_J^{FFF}(x)+C_F \,C_A \, \mathcal{C}_J^{FFA}(x)
+ C_A^2 \, \mathcal{C}_J^{FAA}(x)
\nonumber\\&
+C_F\, T_F\, n_l\, \mathcal{C}_J^{FFL}(x)+ C_F\,  T_F\, n_b\, \mathcal{C}_J^{FFB}(x)+
C_F\,T_F \, n_c\,  \mathcal{C}_J^{FFC}(x)
\nonumber\\&
+C_A\,T_F\, n_l\,\mathcal{C}_J^{FAL}(x) + C_A\,T_F\, n_b\, \mathcal{C}_J^{FAB}(x)  + C_A\,T_F \, n_c\, \mathcal{C}_J^{FAC}(x)
\nonumber\\&
+T_F^2\, n_l^2\, \mathcal{C}_J^{FLL}(x)+T_F^2 \, n_l \, n_b\,  \mathcal{C}_J^{FLB}(x)+T_F^2 \, n_l \, n_c\,  \mathcal{C}_J^{FLC}(x)
\nonumber\\&
+T_F^2\, n_b^2\,
\mathcal{C}_J^{FBB}(x)+ T_F^2 \, n_b \, n_c \, \mathcal{C}_J^{FBC}(x) +
T_F^2 \, n_c^2\,  \mathcal{C}_J^{FCC}(x)
\Big].
\end{align}

In the following, we will present the  numerical results with about 30-digit precision for the
color-structure components of $\mathcal{C}_J^{(2)}(x)$ and $\mathcal{C}_J^{(3)}(x)$ with $J\in\{(t,i0),(t5,ij)\}$ at the physical heavy quark mass ratio  $x=x_0=\frac{150}{475}$:
\begin{align}\label{c23x0}
		\mathcal{C}_{t,i0}^{FF}(x_0) =&-13.7128908053312964335378688241536,
	\nonumber	\\
 \mathcal{C}_{t5,ij}^{FF}(x_0) =&-14.913034700503762441588142929738,
\nonumber	\\
\mathcal{C}_{t,i0}^{FA}(x_0) = \mathcal{C}_{t5,ij}^{FA}(x_0) =&-6.5854991351922034080659088041666 ,
	\nonumber\\
\mathcal{C}_{t,i0}^{FL}(x_0) =\mathcal{C}_{t5,ij}^{FL}(x_0) =& 0.48623749753445268636481818648117,
	\nonumber\\
\mathcal{C}_{t,i0}^{FB}(x_0) =\mathcal{C}_{t5,ij}^{FB}(x_0) =& 0.094767648112565260648796850397580,
	\nonumber\\
\mathcal{C}_{t,i0}^{FC}(x_0) =\mathcal{C}_{t5,ij}^{FC}(x_0) =& 0.58579656372904430515925102361910;
	\nonumber	\\	
\mathcal{C}_{t,i0}^{FFF}(x_0) =& 20.189694171293059999115718422862,
	\nonumber\\	
	\mathcal{C}_{t5,ij}^{FFF}(x_0) =& 22.306062127579275290403925140598,
		\nonumber\\	
\mathcal{C}_{t,i0}^{FFA}(x_0) =& -203.43492648602951942325728768127,
	\nonumber\\
	\mathcal{C}_{t5,ij}^{FFA}(x_0) =& -203.95472214521991932337123118763,
	\nonumber\\
\mathcal{C}_{t,i0}^{FAA}(x_0) =\mathcal{C}_{t5,ij}^{FAA}(x_0) =& -102.79687277377774222247635787879,
	\nonumber\\
\mathcal{C}_{t,i0}^{FFL}(x_0) =&50.937750168903261462489070659559,
	\nonumber\\
	\mathcal{C}_{t5,ij}^{FFL}(x_0) =& 50.848850621112708424855717022108,
	\nonumber\\
	\mathcal{C}_{t,i0}^{FFB}(x_0) =\mathcal{C}_{t5,ij}^{FFB}(x_0) =&-0.12549350490181543572124489903965,
	\nonumber\\
	\mathcal{C}_{t,i0}^{FFC}(x_0) =\mathcal{C}_{t5,ij}^{FFC}(x_0) =& -1.6854789447153670526748653363782 ,
	\nonumber\\
	\mathcal{C}_{t,i0}^{FAL}(x_0) =\mathcal{C}_{t5,ij}^{FAL}(x_0) =& 40.225746623835199555381909178019,
	\nonumber\\
\mathcal{C}_{t,i0}^{FAB}(x_0) = \mathcal{C}_{t5,ij}^{FAB}(x_0) =&-0.20773504228300500317960484318926,
	\nonumber\\
 	\mathcal{C}_{t,i0}^{FAC}(x_0) =	\mathcal{C}_{t5,ij}^{FAC}(x_0) =&0.46466348732388629839619141994117,
 		\nonumber\\
 	\mathcal{C}_{t,i0}^{FLL}(x_0) =\mathcal{C}_{t5,ij}^{FLL}(x_0) =& -2.0881487824796221669234777696960,
	\nonumber\\
\mathcal{C}_{t,i0}^{FLB}(x_0) = \mathcal{C}_{t5,ij}^{FLB}(x_0) =&-0.055625961762816926133354428478288,
	\nonumber\\
\mathcal{C}_{t,i0}^{FLC}(x_0) =\mathcal{C}_{t5,ij}^{FLC}(x_0) =& -0.77633957612352777786750825747681,
	\nonumber\\
\mathcal{C}_{t,i0}^{FBB}(x_0) =\mathcal{C}_{t5,ij}^{FBB}(x_0) =& 0.0155302263395316874159466507909598,
	\nonumber\\
 \mathcal{C}_{t,i0}^{FBC}(x_0) =\mathcal{C}_{t5,ij}^{FBC}(x_0) =&0.090304843884397461649988047441091,
	\nonumber\\
\mathcal{C}_{t,i0}^{FCC}(x_0) =\mathcal{C}_{t5,ij}^{FCC}(x_0) =&0.166410566769625472334622650374377 ,
	\end{align}
where  the color-structure components of  ${\cal C}_{t,i0}^{(n)}(x_0)$ are directly obtained from ${\cal C}_{t,i0}^{(n)}(x)\equiv{\cal C}_{v,i}^{(n)}(x)$  while those of ${\cal C}_{t5,ij}^{(n)}(x_0)$ are calculated by  ${\mathcal{C}_{t5,ij}}=z_t^gz_t^\mu{\overline{\mathcal{C}}_{t5,ij}}$.

We want to mention that all contributions up to N$^3$LO have been calculated for a general QCD gauge parameter $\xi$ ($\xi=0$ corresponds to Feynman gauge) but only with the $\xi^0,\xi^1$ terms, and the final N$^3$LO results of the  matching coefficients   for
the heavy flavor-changing spatial-temporal tensor and spatial-spatial axial-tensor currents are all independent of $\xi$, which constitutes an important check on our calculation.
In the ancillary file, we provide the numerical results with about 30-digit precision for the color-structure components  of $\mathcal{C}_J^{(2)}(x)$ and $\mathcal{C}_J^{(3)}(x)$  at the ten points~\footnote{
	It is worth mentioning that at the point $x=204/498$ the agreement between
	our three-loop numerical results of ${\cal C}_{v,i}$  (${\cal C}_{v,i}\equiv{\cal C}_{t,i0}$) 
	and the corresponding results of the ${\cal C}$ 
	in  eqs.~(20a)--(20o) in ref.~\cite{Sang:2022tnh} is limited to a precision  of only about two
	significant digits.} of $x$ in eq.~\eqref{x10}.

\begin{figure}[htbp]
	\centering
	\includegraphics[width=0.44\textwidth]{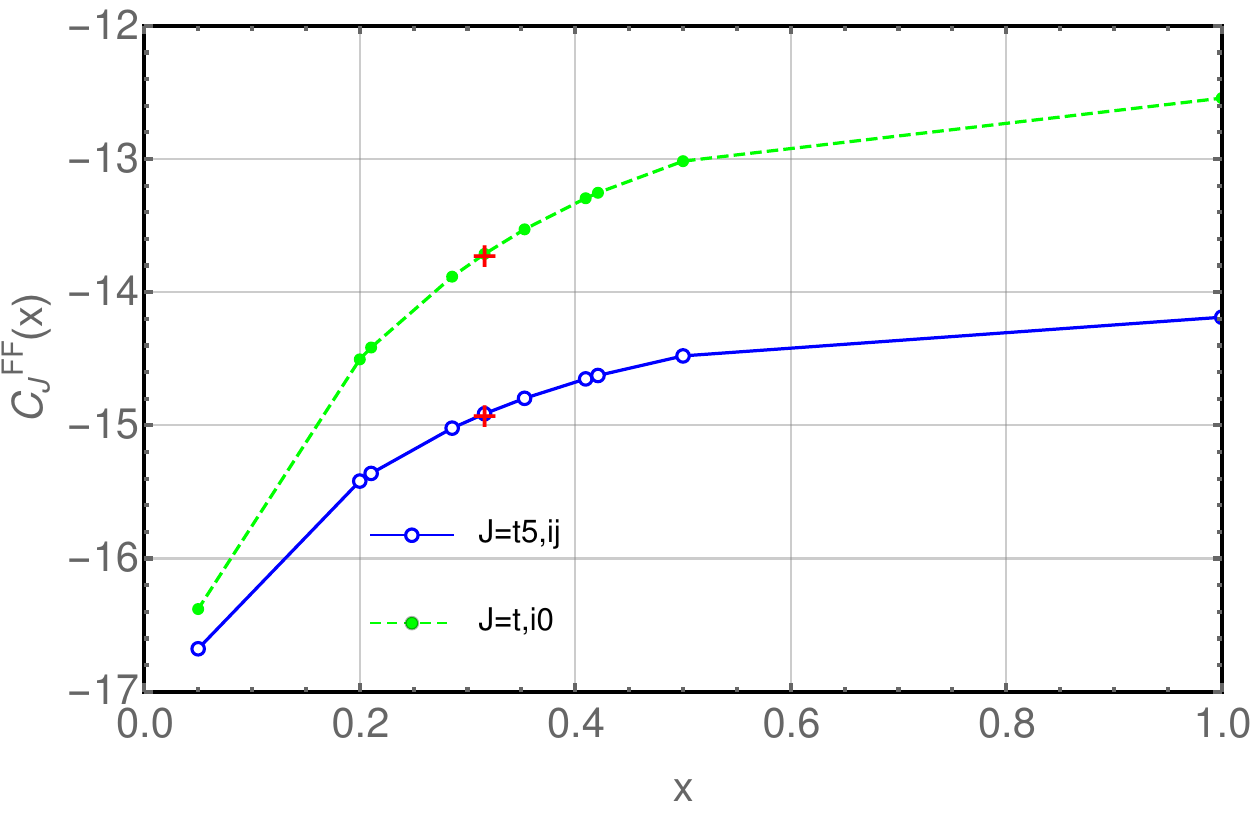}\qquad
	\includegraphics[width=0.44\textwidth]{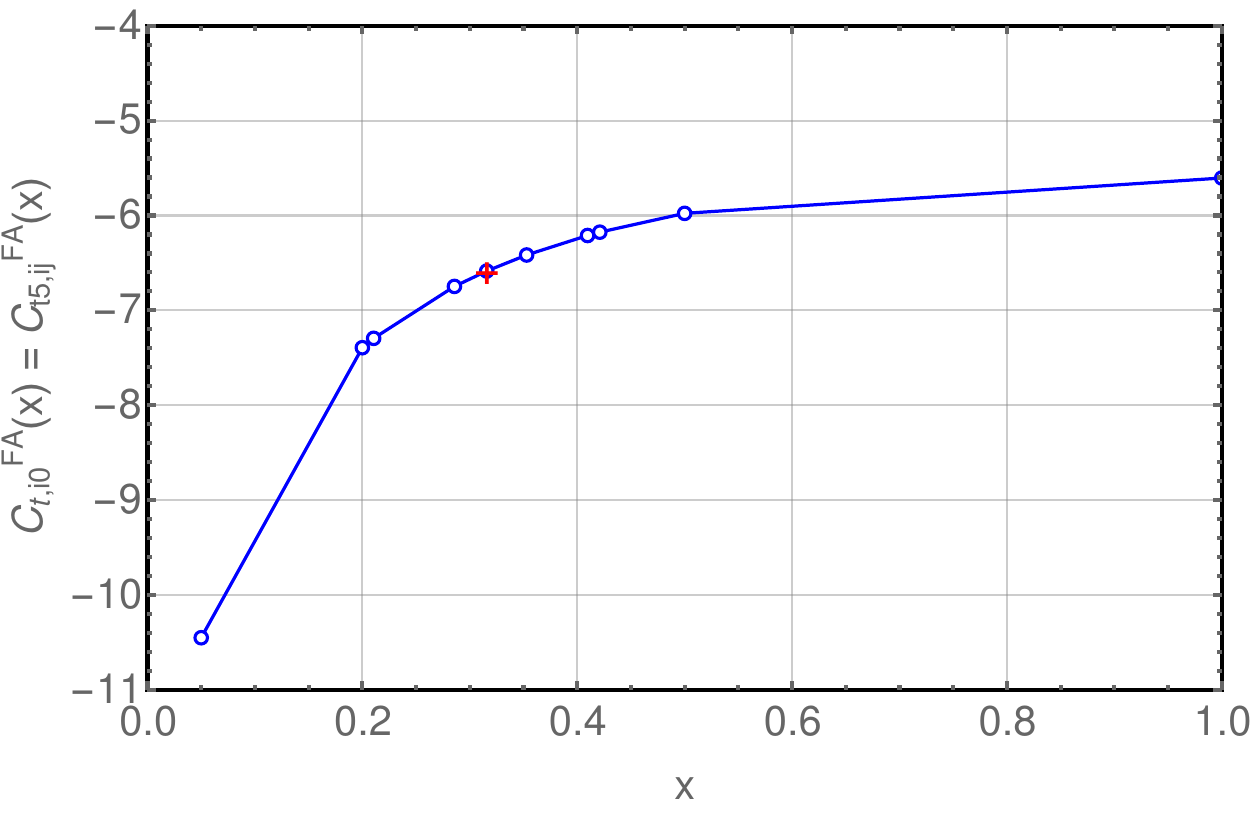}\qquad
	\includegraphics[width=0.44\textwidth]{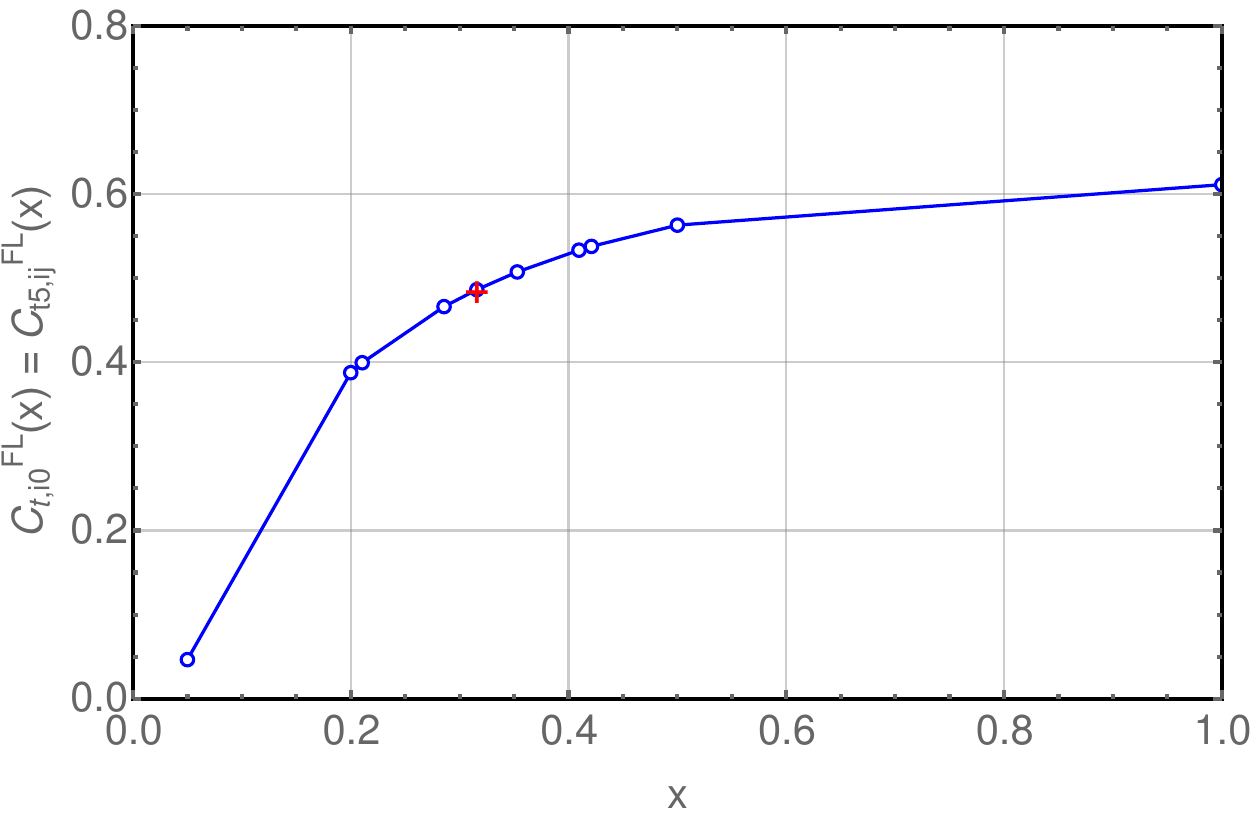}\qquad
	\includegraphics[width=0.44\textwidth]{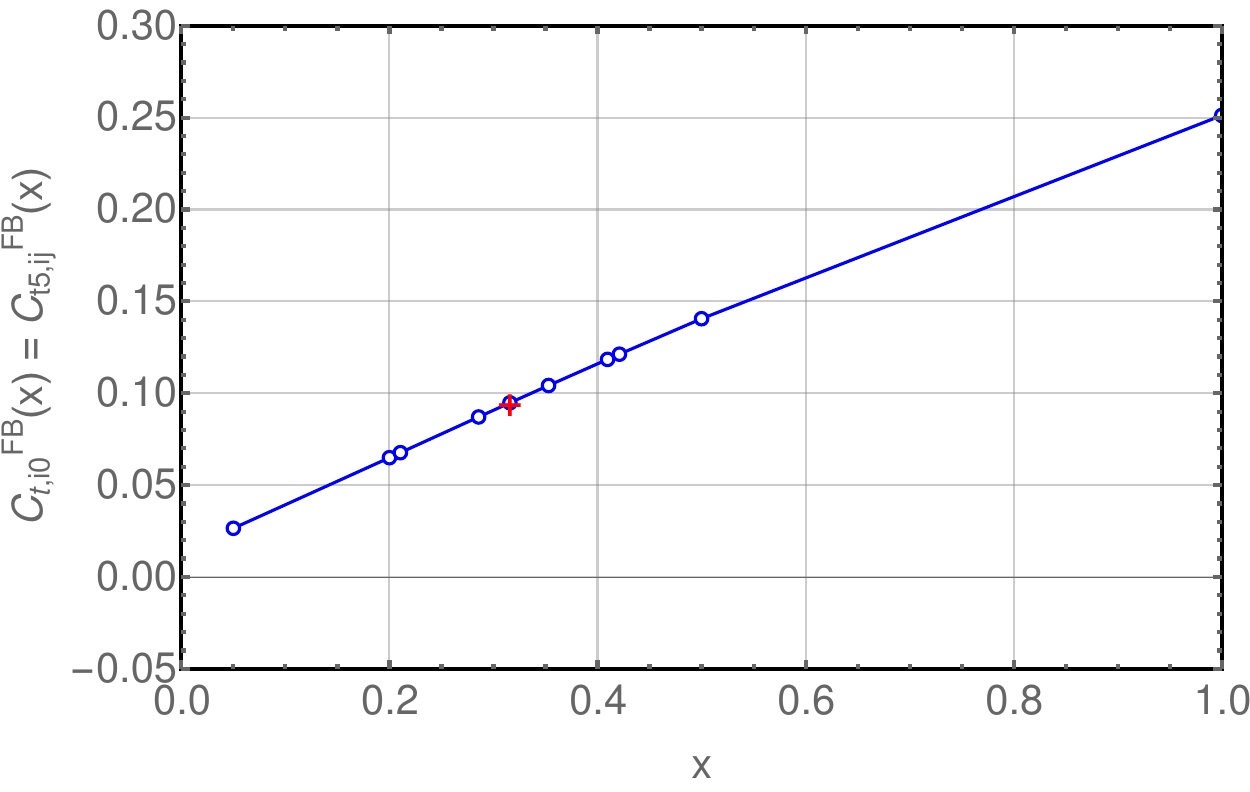}\qquad
	\includegraphics[width=0.44\textwidth]{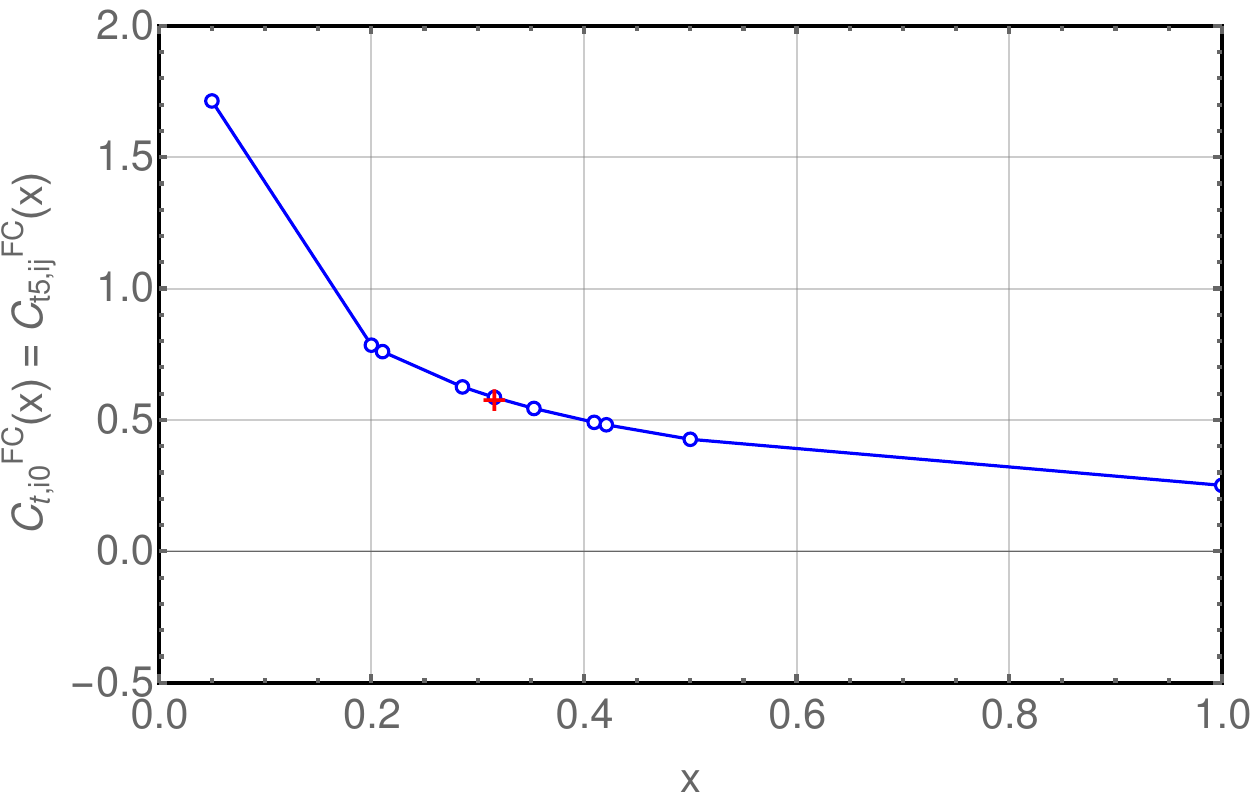}\qquad
		\includegraphics[width=0.44\textwidth]{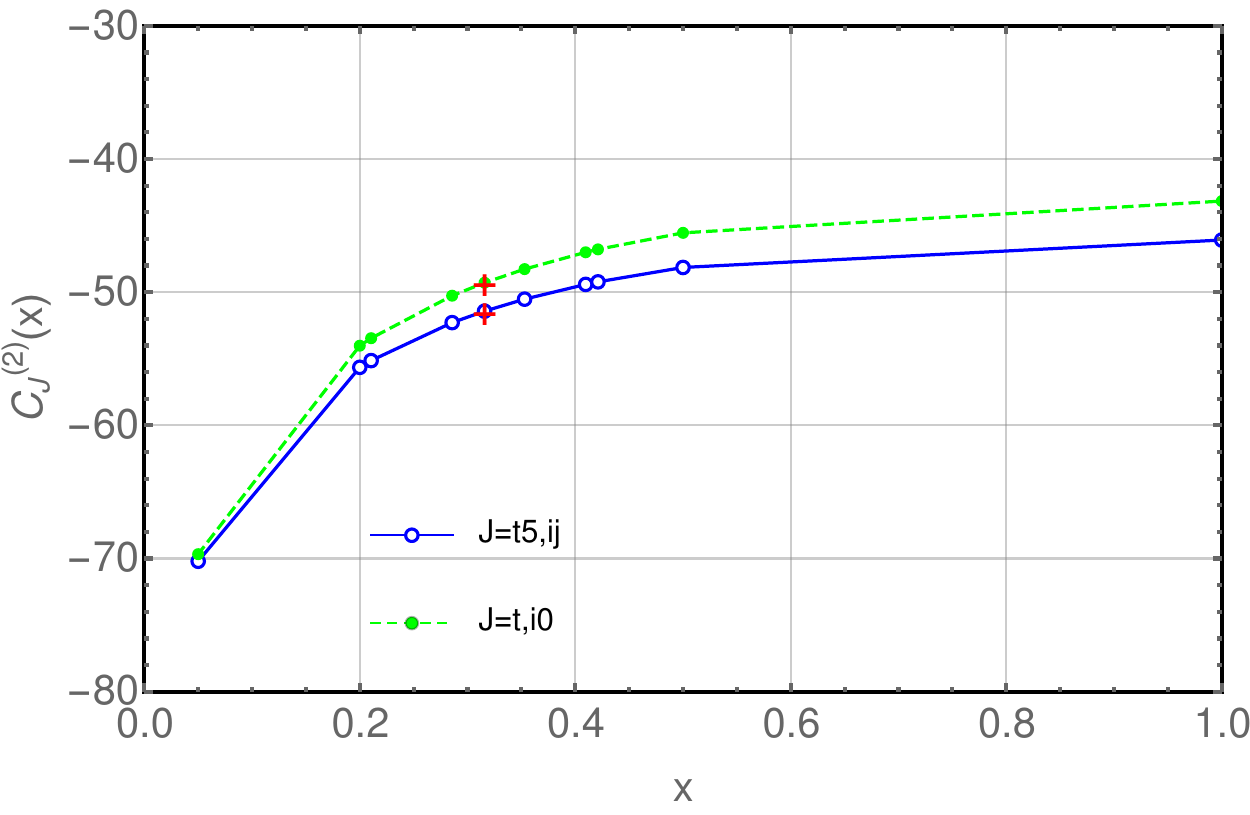}
	\caption{The two-loop coefficient $\mathcal{C}_{J}^{(2)}(x)~(J\in\{(t,i0),(t5,ij)\})$ with $n_l=3,n_b=n_c=1$ and its five color-structure components as functions of the heavy quark mass ratio $x$ within the range of $x\in(0,1]$. The blue hollow dots and green solid dots on the curves represent sample points at ten different values of $x$ in eq.~\eqref{x10}. The red crosses on the curves    correspond to the results at the physical heavy quark mass ratio with $x=x_0=150/475$.}
	\label{fig:c2x}
\end{figure}

\begin{figure}[htbp]
	\centering
	\includegraphics[width=0.315\textwidth]{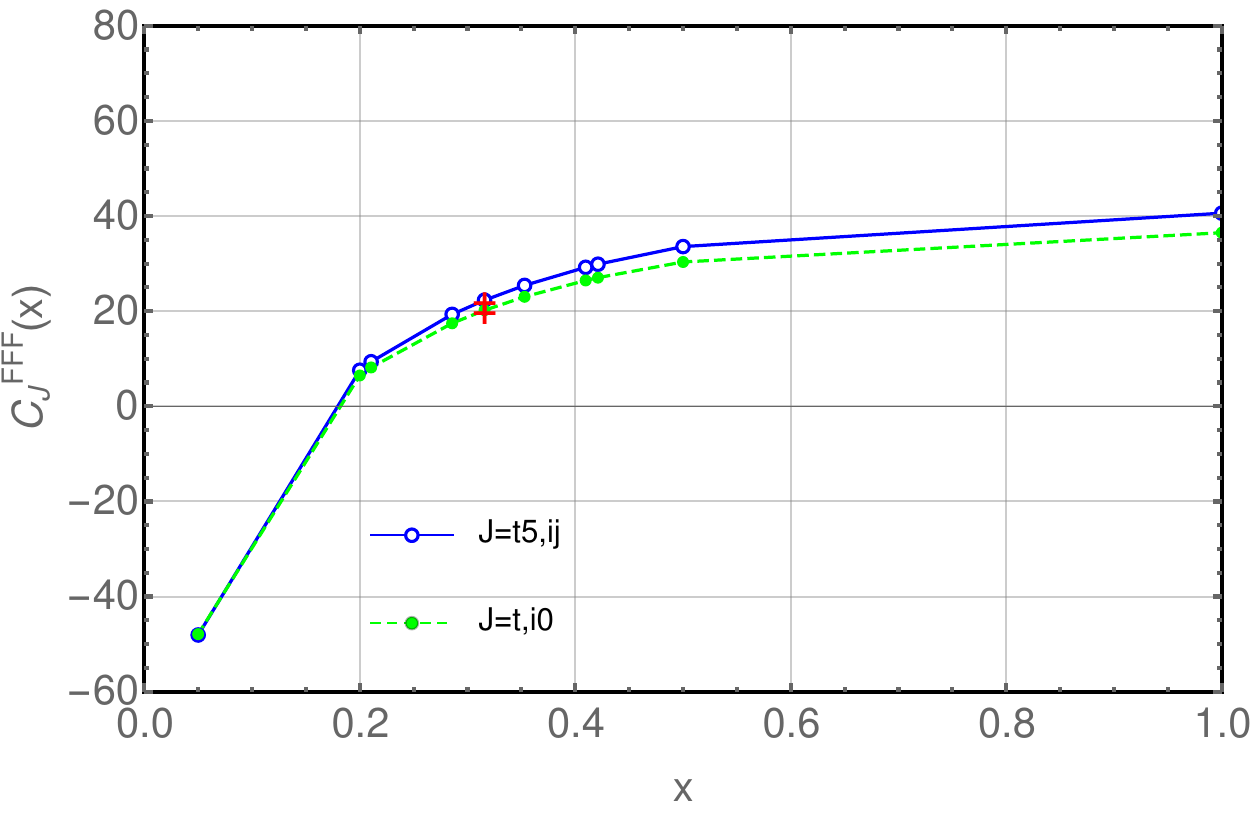}\quad 
	\includegraphics[width=0.315\textwidth]{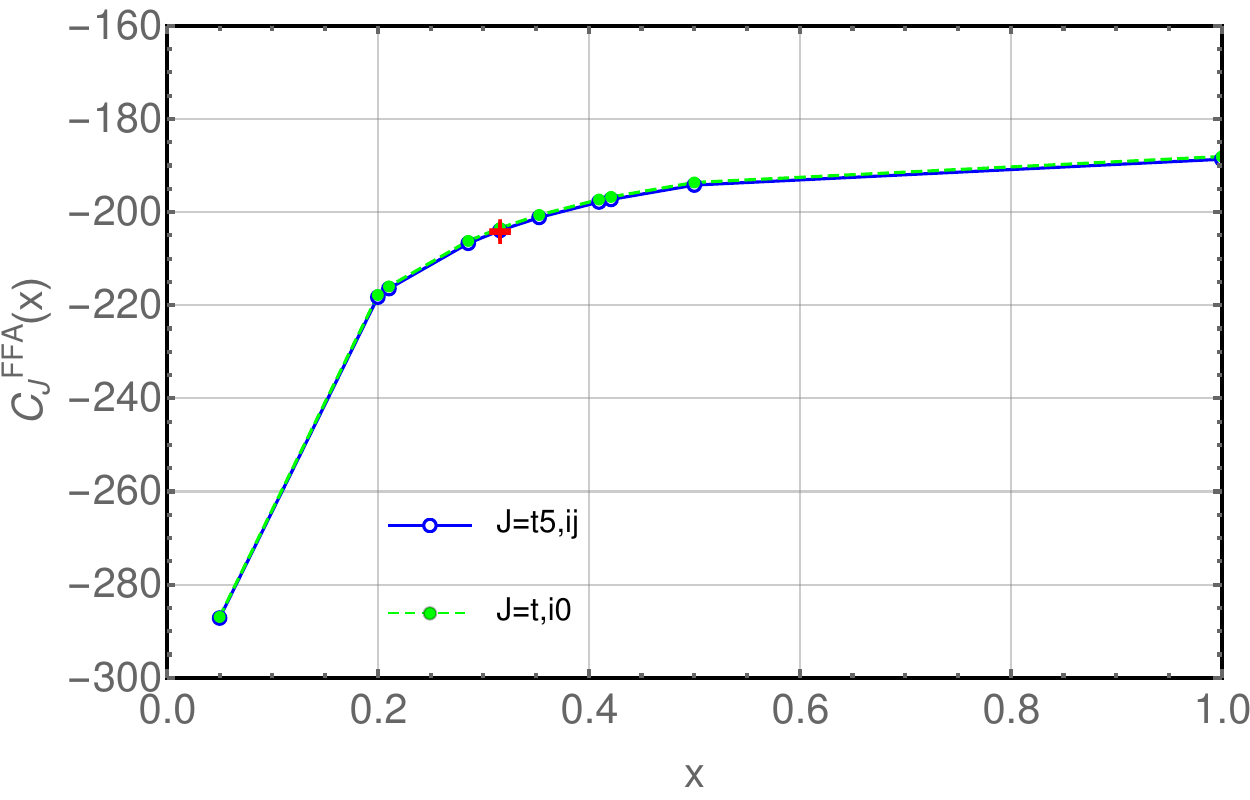}\quad
	\includegraphics[width=0.315\textwidth]{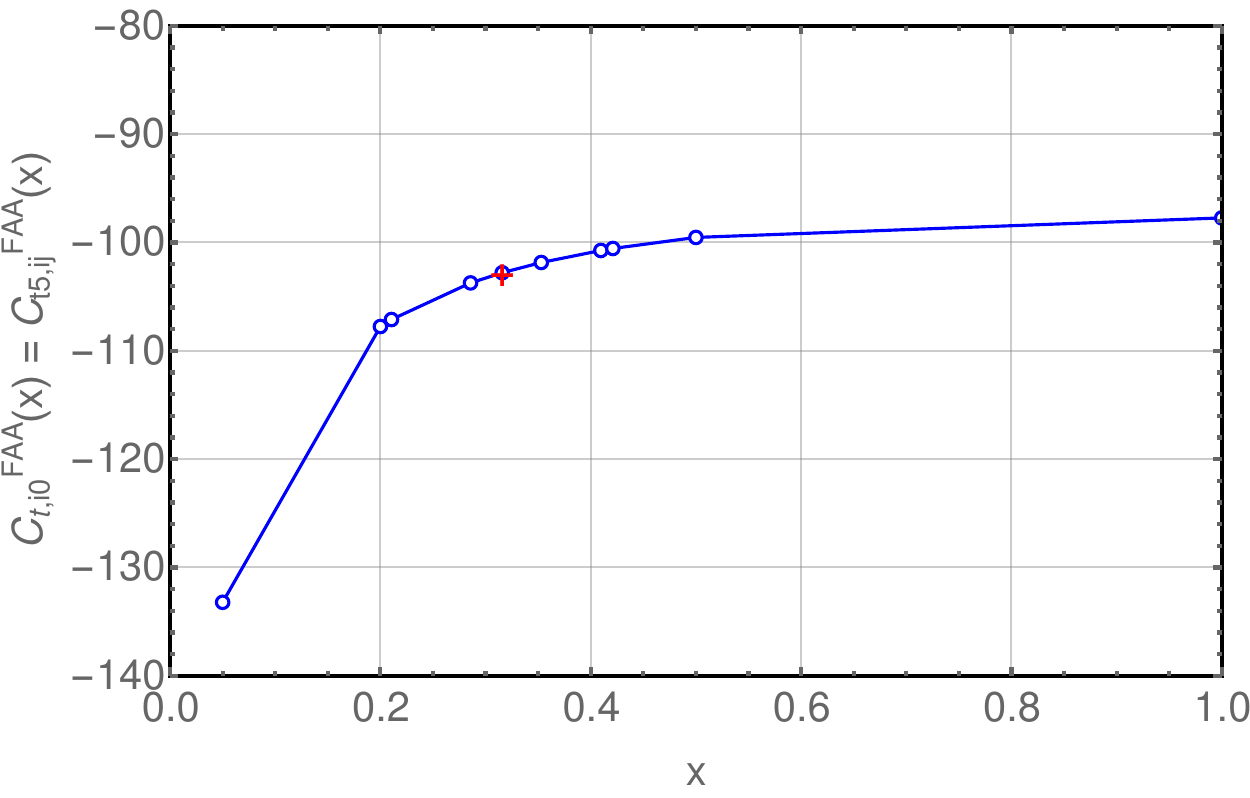}\quad
	\includegraphics[width=0.315\textwidth]{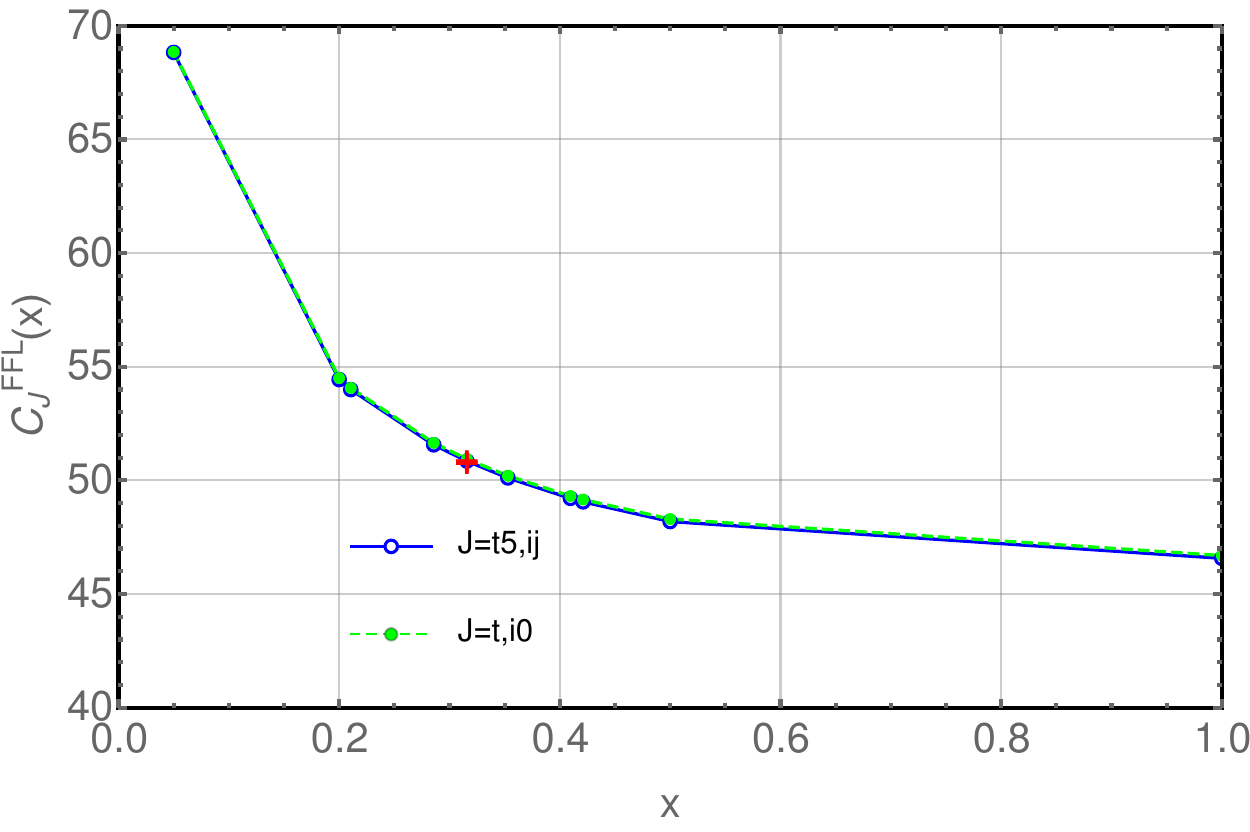}\quad
	\includegraphics[width=0.315\textwidth]{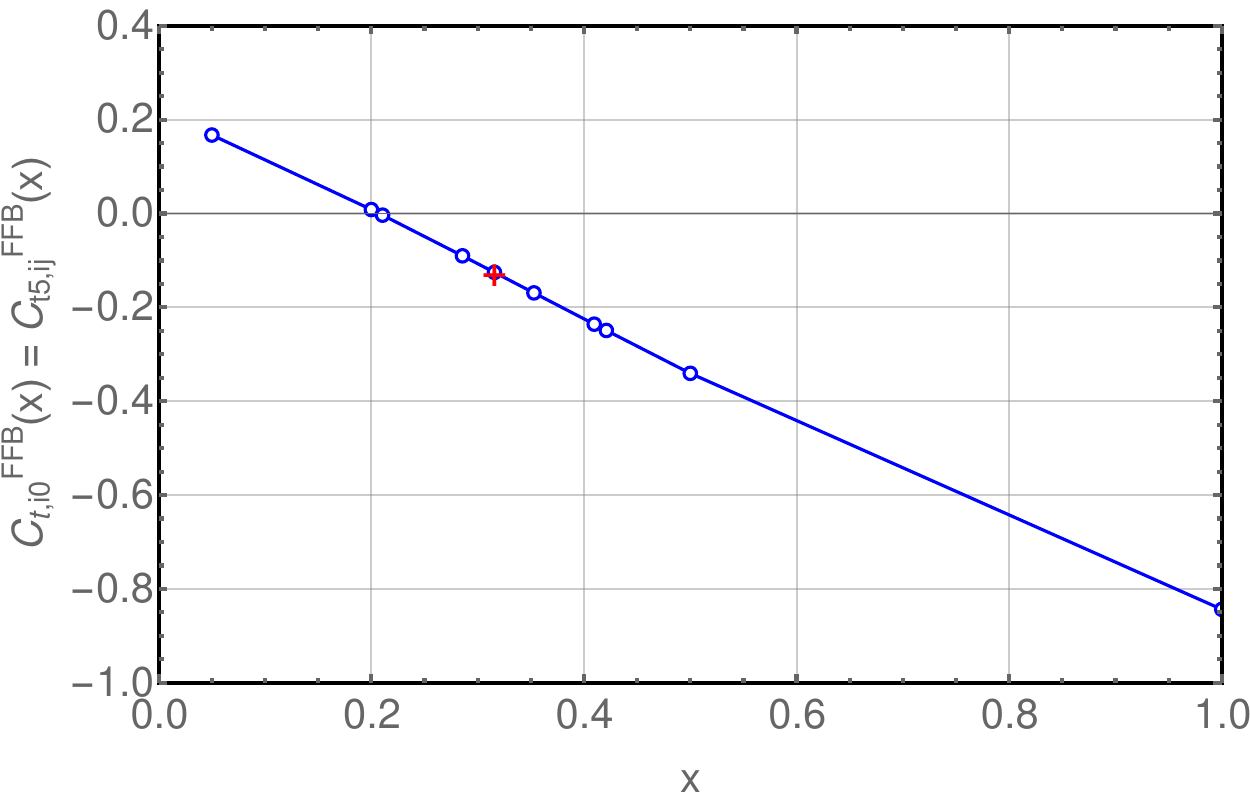}\quad
	\includegraphics[width=0.315\textwidth]{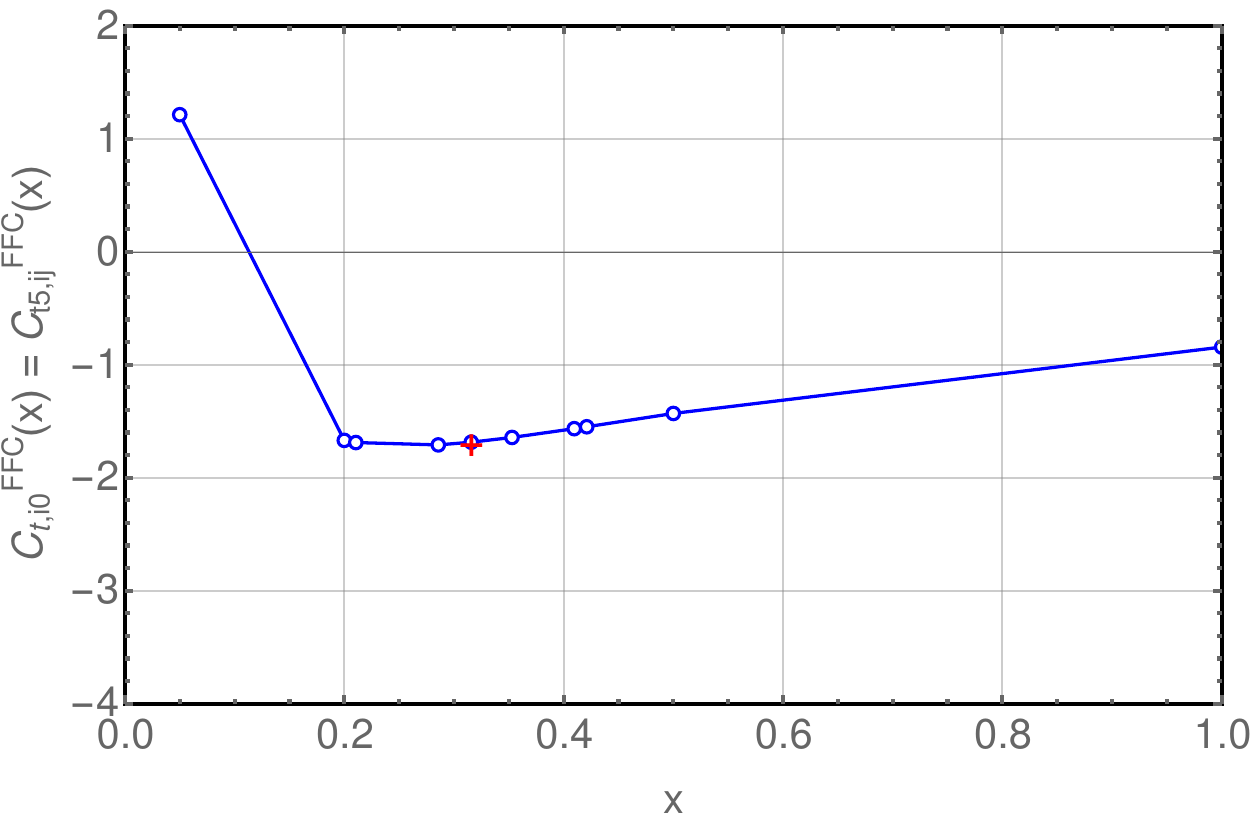}\quad
	\includegraphics[width=0.315\textwidth]{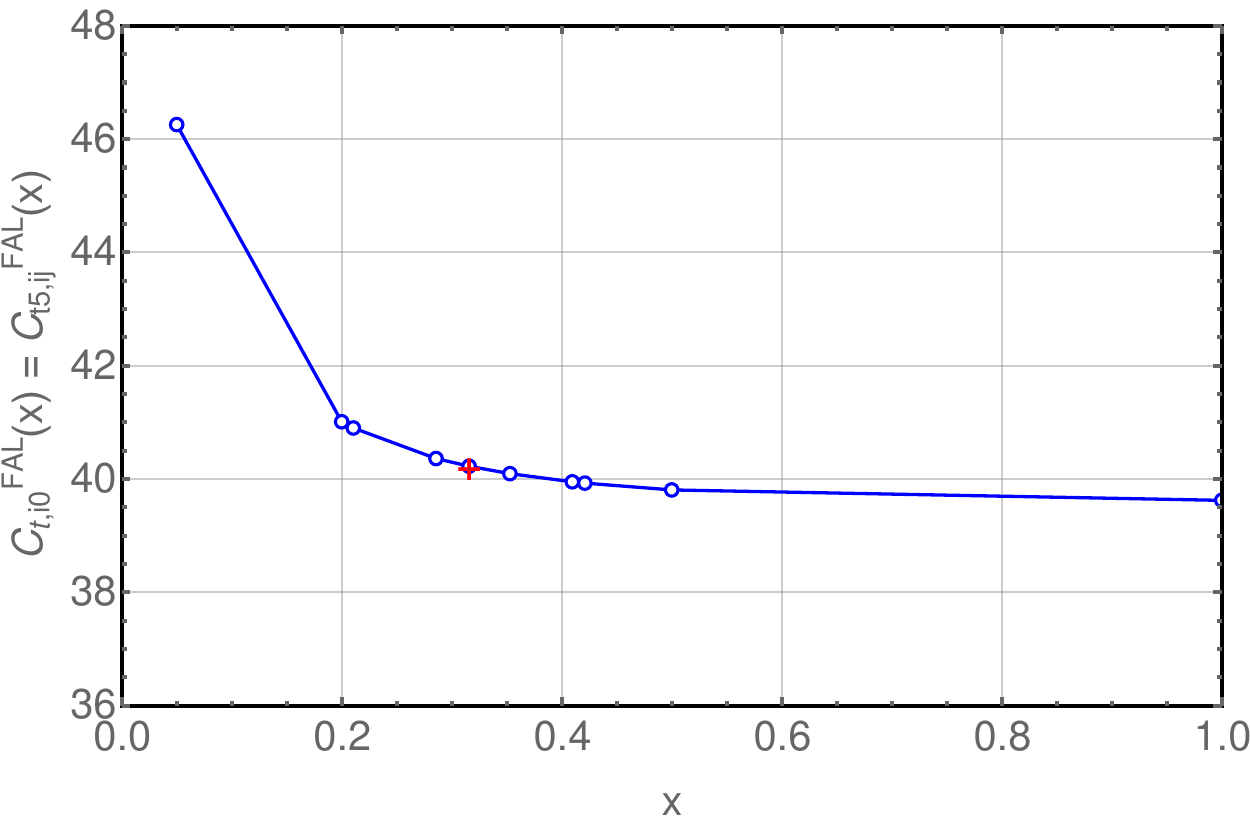}\quad	
	\includegraphics[width=0.315\textwidth]{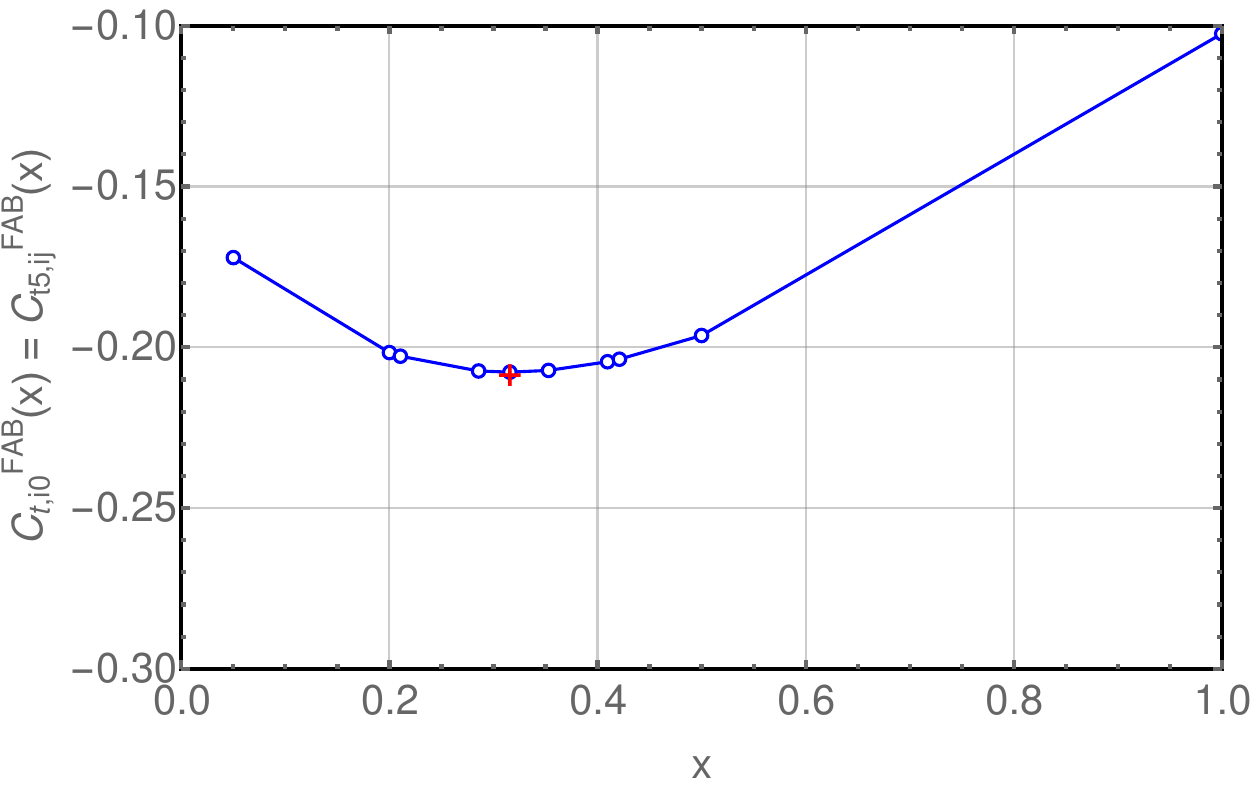}\quad
	\includegraphics[width=0.315\textwidth]{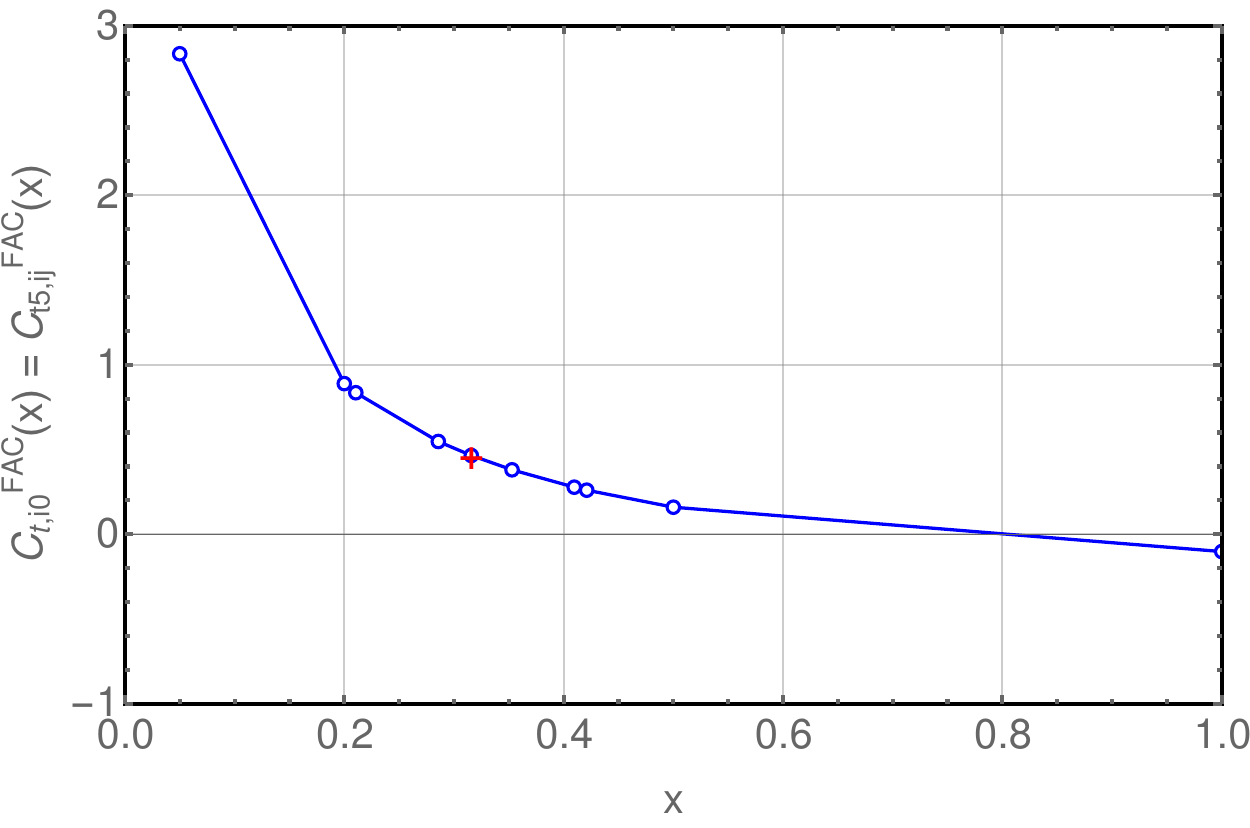}\quad
	\includegraphics[width=0.315\textwidth]{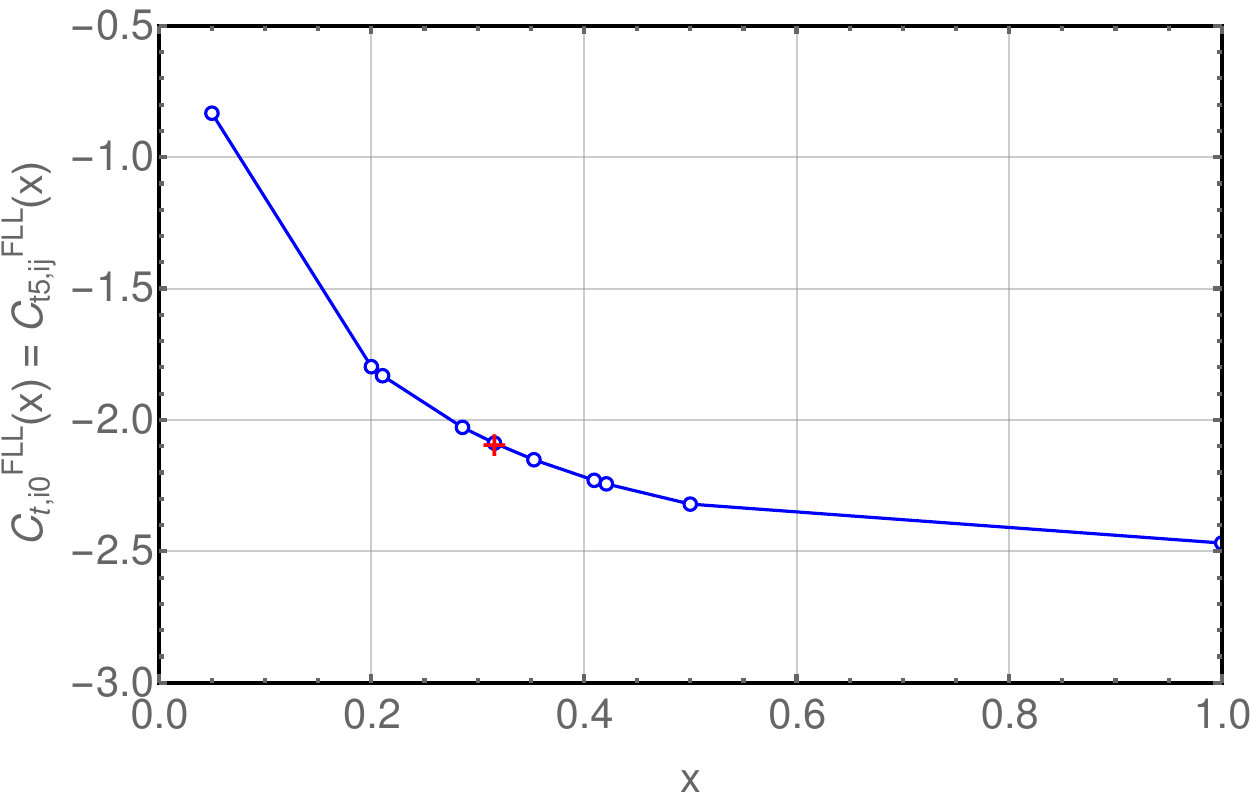}\quad
	\includegraphics[width=0.315\textwidth]{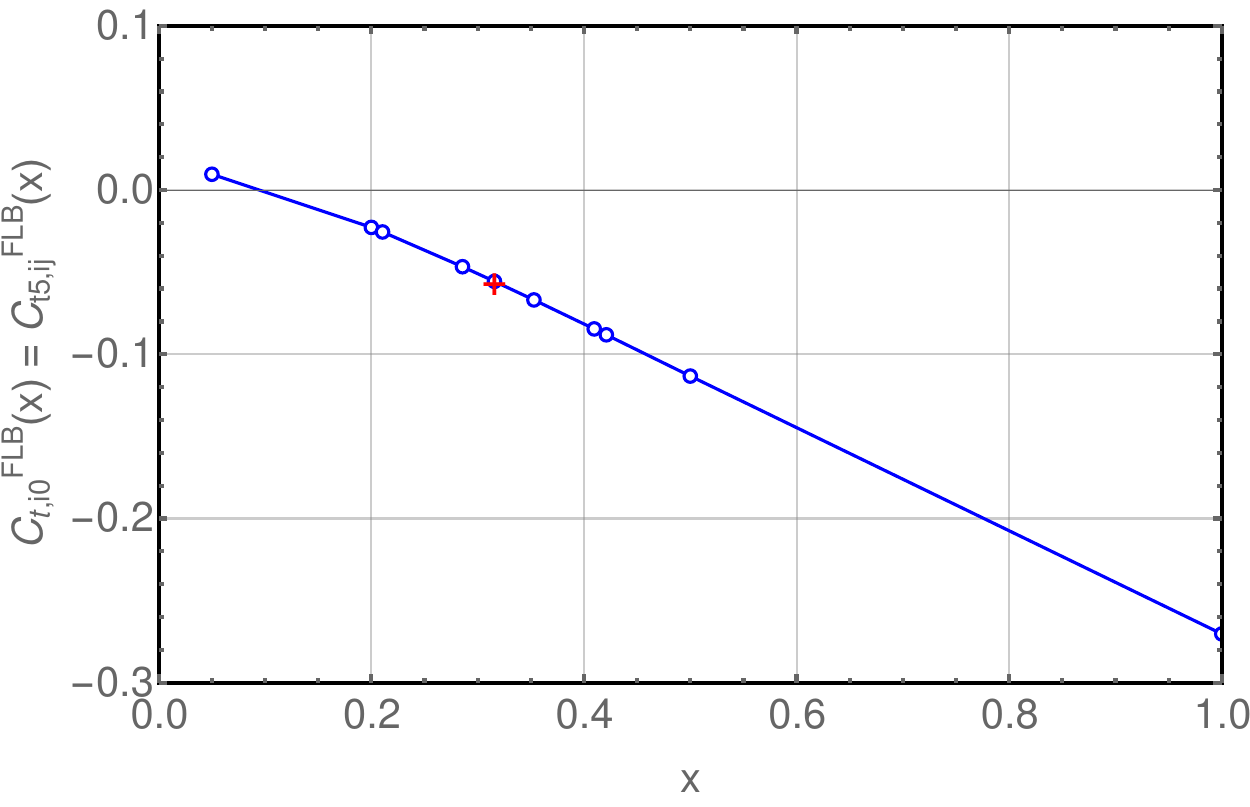}\quad
	\includegraphics[width=0.315\textwidth]{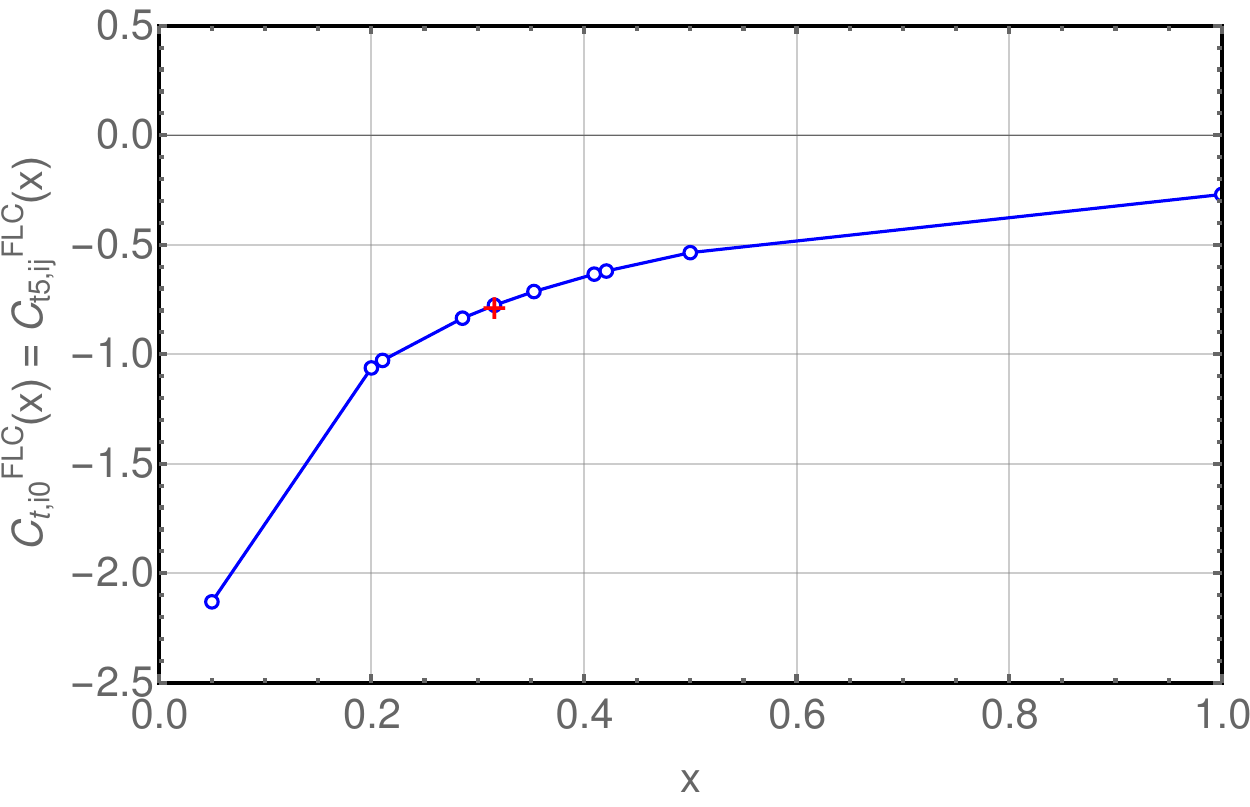}\quad
	\includegraphics[width=0.315\textwidth]{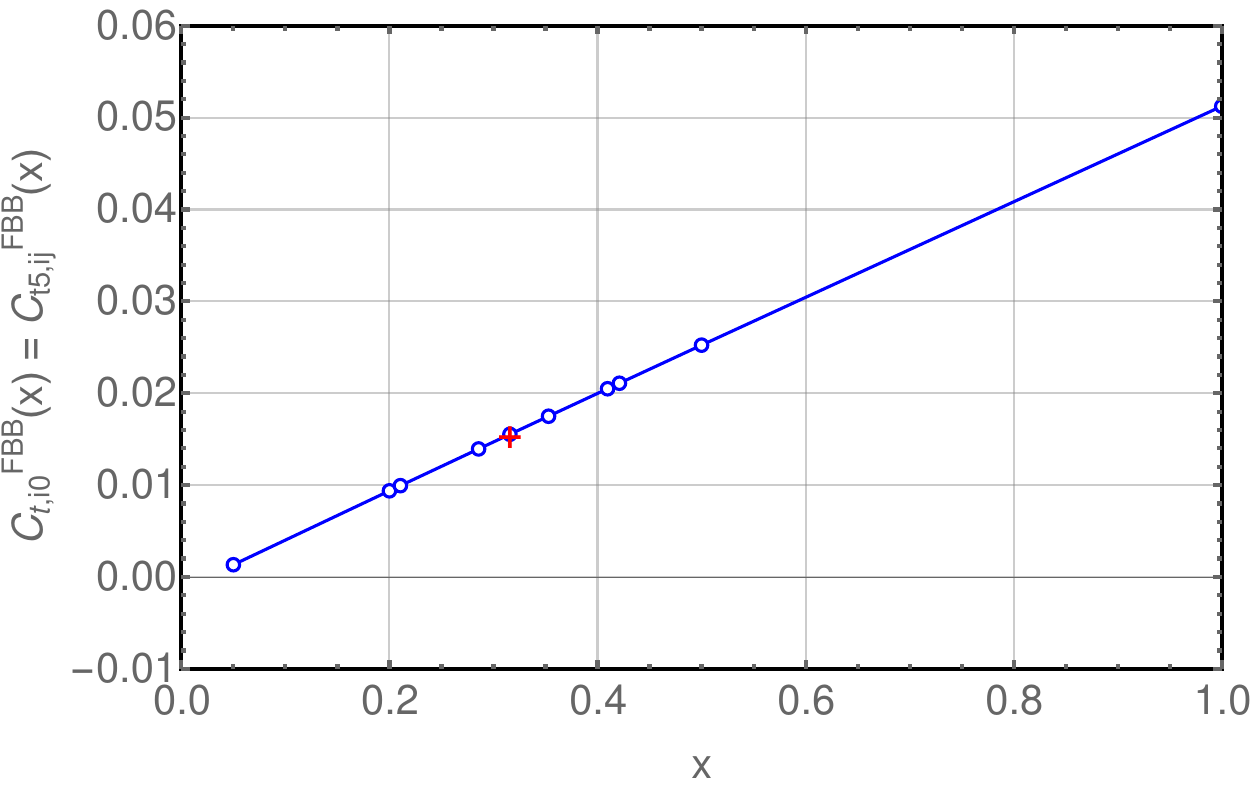}\quad			
	\includegraphics[width=0.315\textwidth]{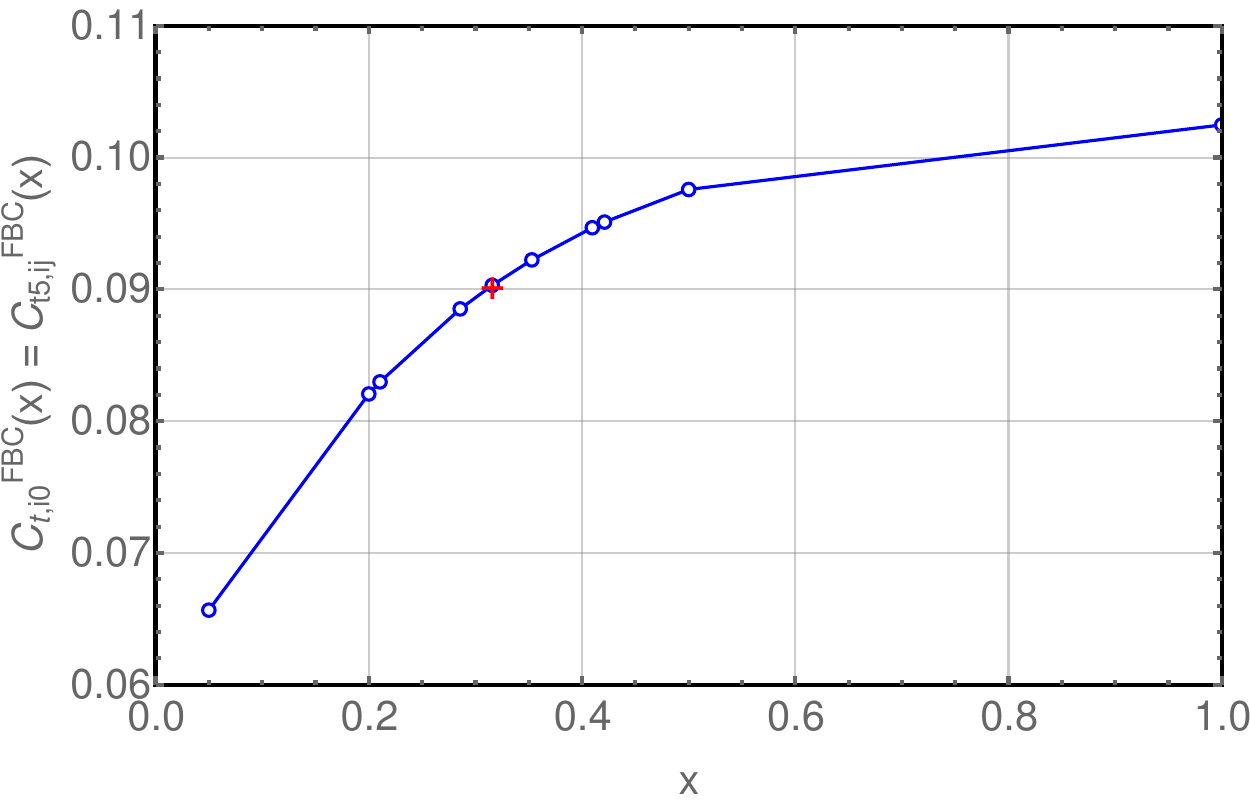}\quad
	\includegraphics[width=0.315\textwidth]{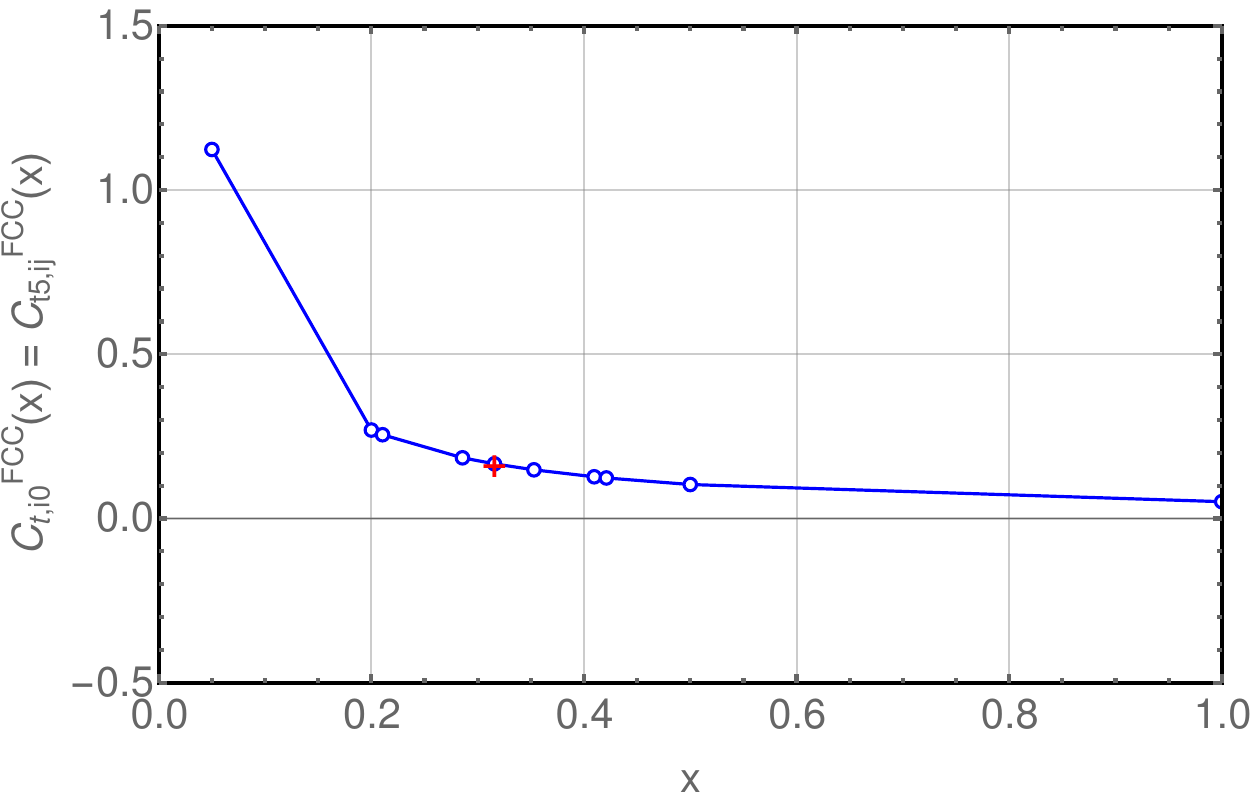}\quad
		\includegraphics[width=0.315\textwidth]{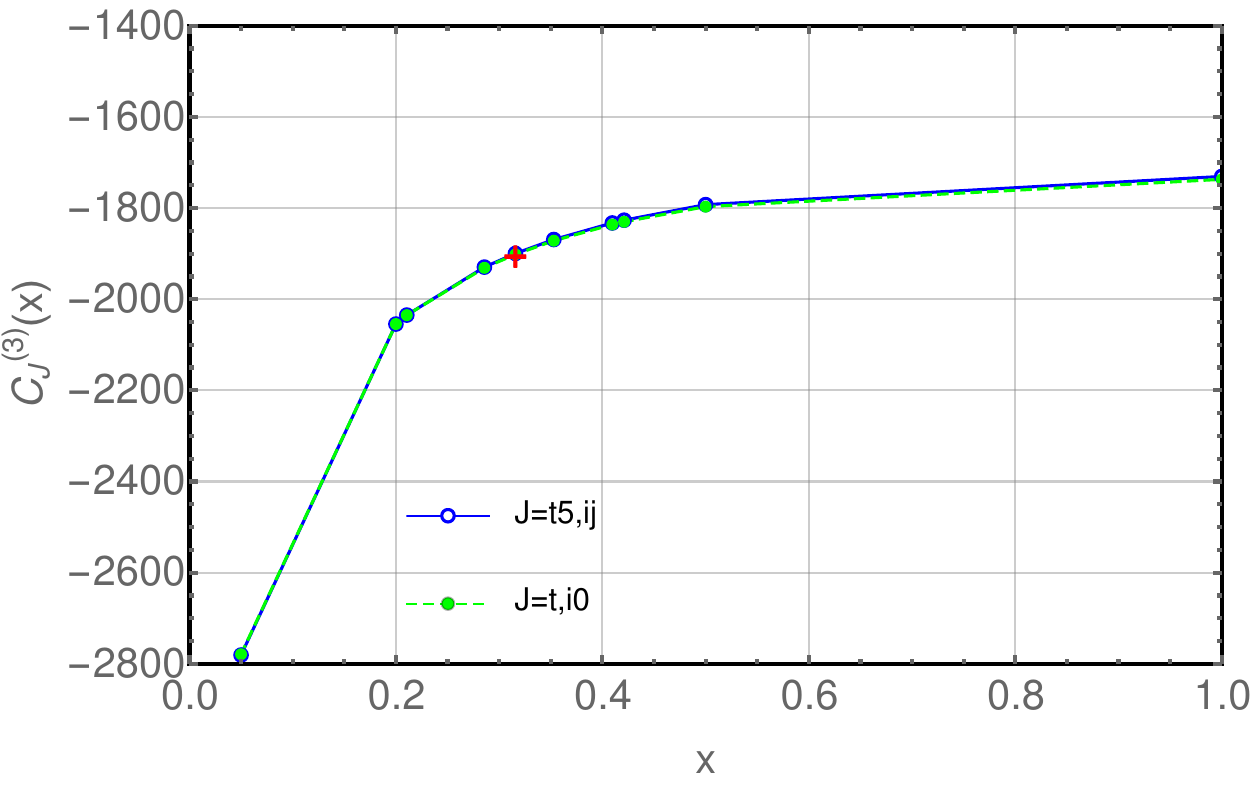}
	\caption{The same as figure~\ref{fig:c2x}, but for the three-loop coefficient $\mathcal{C}_{J}^{(3)}(x)~(J\in\{(t,i0),(t5,ij)\})$ with $n_l=3,n_b=n_c=1$ and its fifteen color-structure components. }
	\label{fig:c3x}
\end{figure}

Choosing our results at the ten points of $x$ as sample data points,
we plot the  dependence of $\mathcal{C}_{J}^{(n)}(x)~(J\in\{(t,i0),(t5,ij)\},n\in\{2,3\})$  with $n_l=3,n_b=n_c=1$ and its  color-structure components on  the	 heavy quark mass ratio $x$   within the range of $x\in(0,1]$
 in figure~\ref{fig:c2x}	and figure~\ref{fig:c3x}, from which one can see $\mathcal{C}_{J}^{(n)}(x)$ and its color-structure components have a relatively weak $x$-dependence in the physical region, indicating that the $B_c^*$ meson might be viewed both as a heavy-heavy meson and as a heavy-light meson~\cite{McNeile:2012qf,Colquhoun:2015oha}.
From eq.~\eqref{c23x0} and  figures~\ref{fig:c2x} and \ref{fig:c3x}, we find the dominant contributions in $\mathcal{C}_J^{(2)}(x)$ and $\mathcal{C}_J^{(3)}(x)$ come from the components corresponding to the color structures $C_F^2$, $C_F C_A$, $C_F^2C_A$ and $C_FC_A^2$, while the contributions from the bottom and charm quark loops are negligible.
We also find almost all color-structure components of ${\cal C}_{t5,ij}^{(n)}(x)$ are  exactly equal to the corresponding components of ${\cal C}_{t,i0}^{(n)}(x)$, except that $|{\cal C}_{t5,ij}^{FFF}(x)-{\cal C}_{t,i0}^{FFF}(x)|\gtrsim|{\cal C}_{t5,ij}^{FF}(x)-{\cal C}_{t,i0}^{FF}(x)|>|{\cal C}_{t5,ij}^{FFA}(x)-{\cal C}_{t,i0}^{FFA}(x)|\gtrsim|{\cal C}_{t5,ij}^{FFL}(x)-{\cal C}_{t,i0}^{FFL}(x)|\gtrsim 0$.

The values of $\mathcal{C}_{J}^{(n)}(x)$ and its color-structure components for $x>1$ can be obtained by employing the invariance~\cite{Tao:2023mtw,Braaten:1995ej,Hwang:1999fc,Lee:2010ts,Onishchenko:2003ui,Chen:2015csa,Feng:2022ruy,Sang:2022tnh} of ${\cal C}_J$  under the exchange $m_b\leftrightarrow m_c$ meanwhile $n_b\leftrightarrow n_c$.
Furthermore, we have checked that both $\mathcal{C}_{J}^{(2)}(x)$ and $\mathcal{C}_{J}^{(3)}(x)$ for $J\in\{(v,i),(t,i0),(t5,ij)\}$ are indeed approximately
linear with respect to $\frac{1}{x}$  in the range of $\frac{1}{x} \in [2,4]$ as the description for the $\mathcal{C}^{(3)}(r)$ in figure~3 in ref.~\cite{Sang:2022tnh}. 
However, it's worth noting that the linear approximation    may not be applicable to other values of $x$ within the range of $x \in (0,\infty)$.

We consider the ratio  of the $B_c^*$ decay constant involving the spatial-temporal tensor current to that involving the spatial-spatial axial-tensor current, from which the wave function at the origin is eliminated~\cite{Beneke:2014qea,Rauh:2018vsv,Tao:2023mtw,Beneke:1997jm,Onishchenko:2003ui,Sang:2022tnh,Feng:2022ruy,Sang:2023cwn} so that the ratio of the physical decay constants is approximately equal to the ratio of the nonphysical matching coefficients~\cite{Broadhurst:1994se}, i.e.
\begin{align}\label{ratiof}
\frac{f_{B_c^*}^{t,i0}}{f_{B_c^*}^{t5,ij}}\approx\frac{\mathcal{C}_{t,i0} \times |\Psi_{B_c^*}(0)|}{\mathcal{C}_{t5,ij} \times |\Psi_{B_c^*}(0)|}\approx\frac{\mathcal{C}_{t,i0}}{\mathcal{C}_{t5,ij}}.
\end{align}

Throughout our calculation in the remaining part of this section,   we will expand both the matching coefficients and the ratio of the matching coefficients (decay constants) in power series of ${\alpha_s^{(n_l=3)}(\mu)}$ and study the numerical results up to $\mathcal{O}(\alpha_s^3)$ for them.
Setting $\mu_f=1.2\,\mathrm{GeV}$, $\mu=\mu_0=3\mathrm{GeV}$, $m_b=4.75\mathrm{GeV}$, $m_c=1.5\mathrm{GeV}$,  the $\alpha_s$-expansions of eq.~\eqref{Cjformula} and eq.~\eqref{ratiof}    reduce to
\begin{align}\label{asexpandexprnum}
\mathcal{C}_{t,i0}
=&1 - 2.067273 \frac{\alpha_s^{(3)}(\mu_0)}{\pi} -
29.29166 \left(\frac{\alpha_s^{(3)}(\mu_0)}{\pi}\right)^2 -
1689.867 \left(\frac{\alpha_s^{(3)}(\mu_0)}{\pi}\right)^3+\mathcal{O}(\alpha_s^4),
\nonumber\\
\mathcal{C}_{t5,ij}  =&1 - 2.067273 \frac{\alpha_s^{(3)}(\mu_0)}{\pi} -
31.42525 \left(\frac{\alpha_s^{(3)}(\mu_0)}{\pi}\right)^2 -
1696.499 \left(\frac{\alpha_s^{(3)}(\mu_0)}{\pi}\right)^3+\mathcal{O}(\alpha_s^4),
\nonumber\\
\frac{f_{B_{c}^*}^{t,i0}}{f_{B_{c}^*}^{t5,ij}}\approx&\frac{\mathcal{C}_{t,i0}}{\mathcal{C}_{t5,ij}}  =1 + 2.133589 \left(\frac{\alpha_s^{(3)}(\mu_0)}{\pi}\right)^2 +
11.04305 \left(\frac{\alpha_s^{(3)}(\mu_0)}{\pi}\right)^3+\mathcal{O}(\alpha_s^4).
\end{align}

\begin{figure}[htbp]
	\centering
	\includegraphics[width=0.44\textwidth]{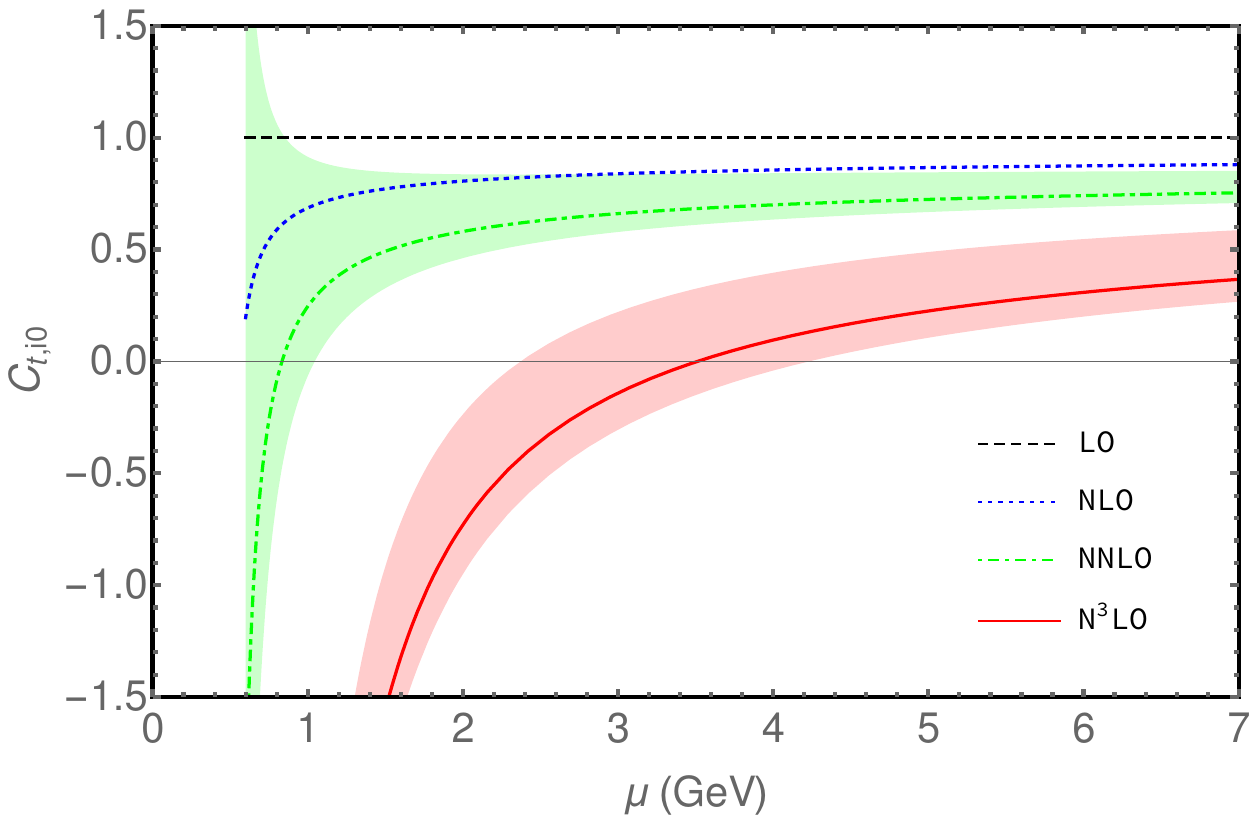}\qquad
	\includegraphics[width=0.44\textwidth]{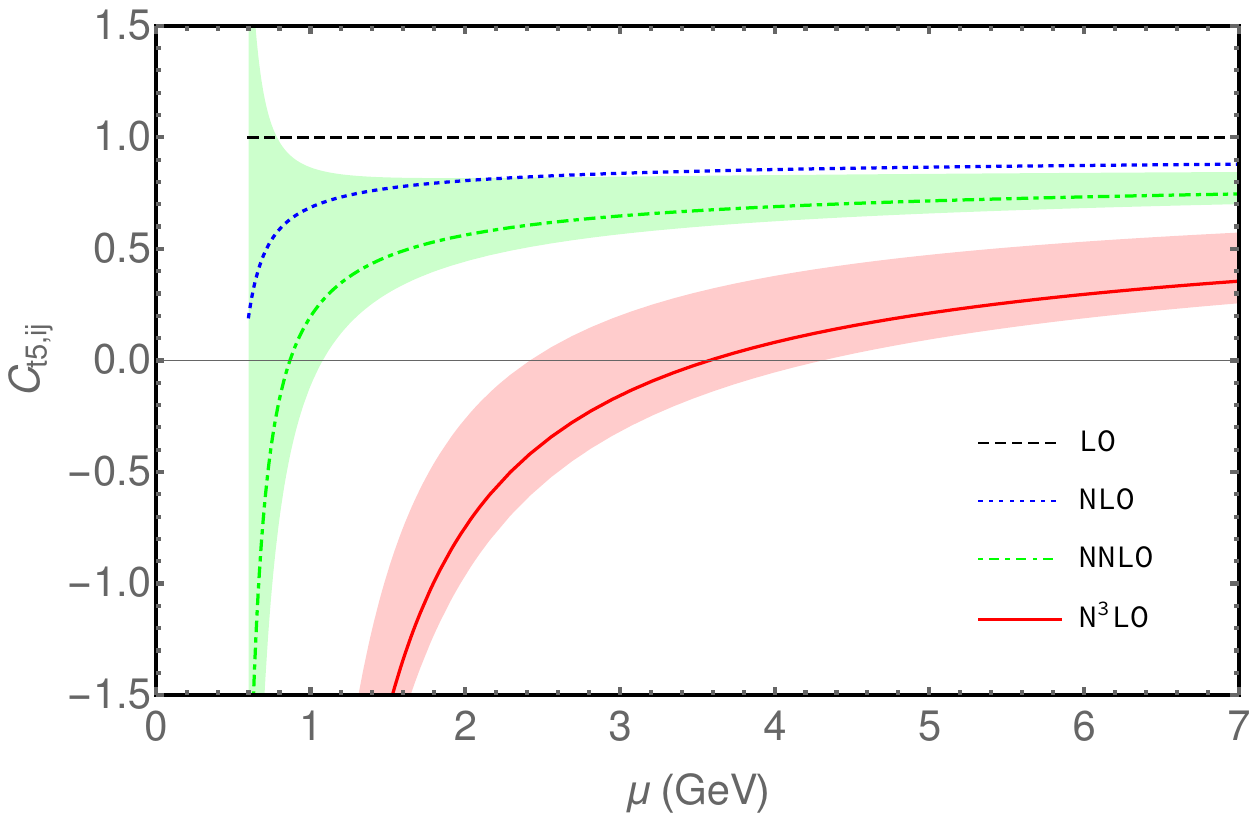}\qquad
	\caption{The renormalization scale $\mu$ dependence of the matching coefficients $\mathcal{C}_{t,i0}$ and $\mathcal{C}_{t5,ij}$
		at LO,  NLO,  NNLO and N$^3$LO accuracy. The central values of  the matching coefficients  are calculated inputting the  physical values with $\mu_f=1.2~\,\mathrm{GeV}$,  $m_b=4.75\mathrm{GeV}$ and $m_c=1.5\mathrm{GeV}$.   The error bands come from the variation of  $\mu_f$  between   2 and 0.4 $\mathrm{GeV}$. }
	\label{fig:Cjmu}
\end{figure}
\begin{figure}[htbp]
	\centering
	\includegraphics[width=0.6\textwidth]{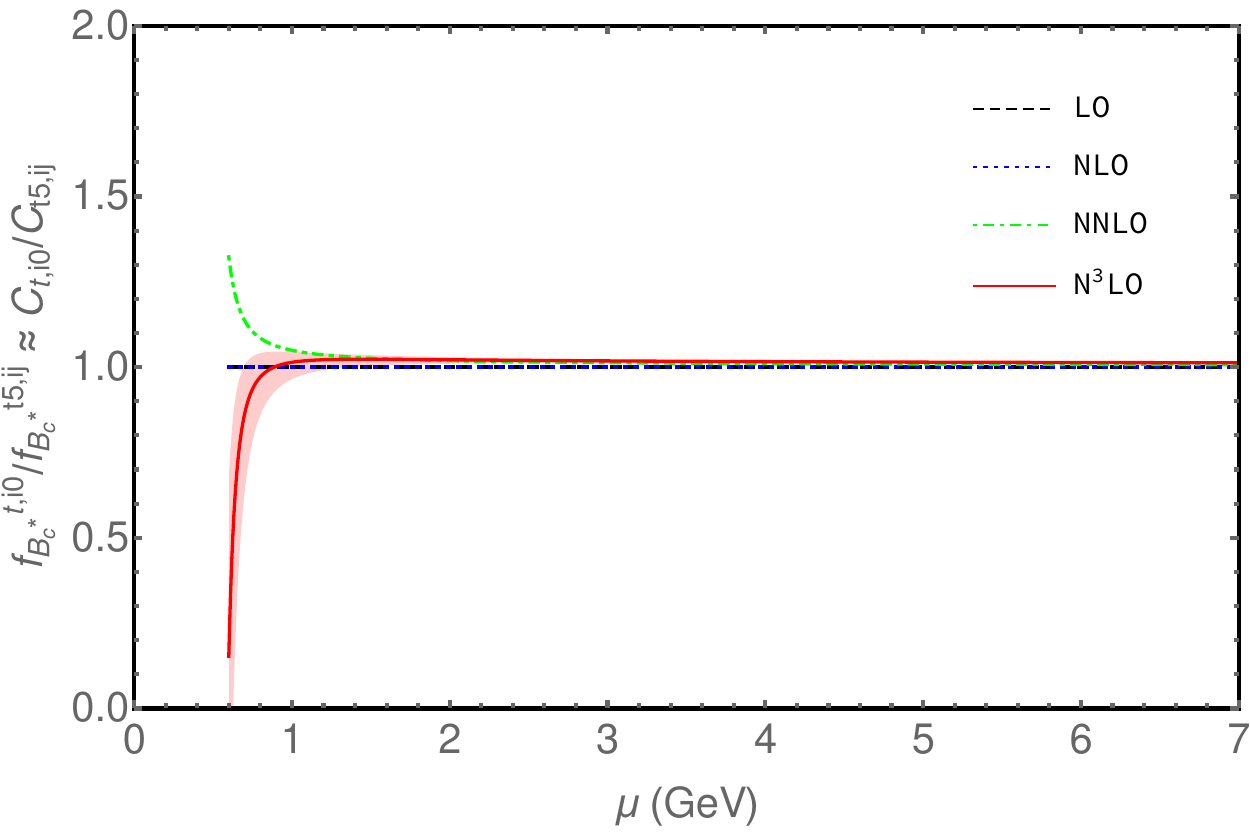}\qquad
	\caption{The renormalization scale $\mu$ dependence of the matching coefficient (decay constant) ratio
		at LO,  NLO,  NNLO and N$^3$LO accuracy. The central values  are calculated inputting the  physical values with  $\mu_f=1.2~\,\mathrm{GeV}$,  $m_b=4.75\mathrm{GeV}$ and $m_c=1.5\mathrm{GeV}$.   The error band comes from the variation of  $\mu_f$  between  7 and 0.4 $\mathrm{GeV}$. }
	\label{fig:CivCjmu}
\end{figure}

\begin{table}[htbp]
	\begin{center}
		\caption{The values of the matching coefficients $\mathcal{C}_{t,i0}$ and $\mathcal{C}_{t5,ij}$   up to N$^3$LO.
			The central values of  the matching coefficients  are calculated inputting the  physical values with $\mu_f=1.2~\mathrm{GeV}$, $\mu=\mu_0=3\mathrm{GeV}$, $m_b=4.75\mathrm{GeV}$ and $m_c=1.5\mathrm{GeV}$.
			The uncertainties are estimated by varying $\mu_f$  from   2 to 0.4 $\mathrm{GeV}$, $\mu$  from   7 to 1.5 $\mathrm{GeV}$, $m_b$  from   5.25 to 4.25 $\mathrm{GeV}$,  $m_c$  from   2 to 1 $\mathrm{GeV}$, respectively.}
		\label{tab:Cjnum}
		\renewcommand\arraystretch{2}
		\tabcolsep=0.20cm
  \resizebox{\linewidth}{!}{
		\begin{tabular}{ c c c c c}
			\hline\hline
			& LO         &  NLO                   & NNLO     & N$^3$LO
			\\  \hline
			$\mathcal{C}_{t,i0}$	 & $1$ & $0.83875^{-0+0.04086+0.00753-0.01927}_{+0-0.06738-0.00790+0.03251}$   &    $0.66053^{-0.08198+0.09301+0.01324+0.02563}_{+0.17632-0.16857-0.01477-0.03155}$  &     $-0.14143^{-0.16360+0.50840+0.01790+0.08745}_{+0.36305-1.41702-0.02078-0.15050}$
			\\  \hline
			$\mathcal{C}_{t5,ij}$	 & $1$ & $0.83875^{-0+0.04086+0.00753-0.01927}_{+0-0.06738-0.00790+0.03251}$   &    $0.64755^{-0.08198+0.09875+0.01392+0.02378}_{+0.17632-0.18168-0.01552-0.02879}$  &     $-0.15756^{-0.16166+0.51277+0.01881+0.08660}_{+0.35888-1.41796-0.02179-0.14861}$	
			\\		\hline \hline
		\end{tabular}}
	\end{center}
\end{table}
\begin{table}[htbp]
	\begin{center}
		\caption{The same as table~\ref{tab:Cjnum}, but for the ratio of  the matching coefficients (decay constants),  with the uncertainties in the first column estimated by  varying $\mu_f$  from   7 to 0.4 $\mathrm{GeV}$.}
		\label{tab:CivCjnum}
		\renewcommand\arraystretch{2}
		\tabcolsep=0.2cm
  \resizebox{\linewidth}{!}{
		\begin{tabular}{ c c c c c}
			\hline\hline
			& LO         &  NLO                   & NNLO     & N$^3$LO	
			\\  \hline
			$\frac{f_{B_{c}^*}^{t,i0}}{f_{B_{c}^*}^{t5,ij}}\approx\frac{\mathcal{C}_{t,i0}}{\mathcal{C}_{t5,ij}}$	 & $1$ & $1$   &    $1.01298^{-0-0.00575-0.00068+0.00186}_{+0+0.01312+0.00074-0.00276}$  &     $1.01822^{-0.00670-0.00559-0.00111+0.00143}_{+0.00417+0.00481+0.00124-0.00267}$
			\\		\hline \hline
		\end{tabular}}
	\end{center}
\end{table}

With the values of $\alpha_s^{\left(n_l=3\right)}(\mu)$ calculated (see Sec.~\ref{ZjNRQCD}),
we investigate the QCD renormalization scale $\mu$ dependence of the matching coefficients and the matching coefficient (decay constant) ratio
at LO,  NLO,  NNLO and N$^3$LO accuracy in figure~\ref{fig:Cjmu} and figure~\ref{fig:CivCjmu}, respectively.
The middle lines correspond to the choice of $\mu_f=1.2\,\mathrm{GeV}$ for the NRQCD factorization scale, and the upper and lower  edges   of the error bands
correspond to $\mu_f=0.4\,\mathrm{GeV}$ and $\mu_f=2(7)~\mathrm{GeV}$, respectively.
Furthermore, we  present our precise numerical results of the matching coefficients and the  matching coefficient (decay constant) ratio  at LO,  NLO,  NNLO and N$^3$LO accuracy   in table~\ref{tab:Cjnum} and table~\ref{tab:CivCjnum}, respectively, where the uncertainties from $\mu_f$, $\mu$, $m_b$ and $m_c$ are included.

From eq.~\eqref{asexpandexprnum}, the figures~\ref{fig:Cjmu} and \ref{fig:CivCjmu}, as well as the tables~\ref{tab:Cjnum} and \ref{tab:CivCjnum}, we have the following points:
\begin{enumerate}
\item[(1)]
Both the matching coefficients ${\cal C}_{t,i0}$ and ${\cal C}_{t5,ij}$ are nonconvergent up to N$^3$LO; especially, the 
third order corrections to them are very large.
Besides, the N$^3$LO corrections to the  matching coefficients also exhibit very strong dependence on both the QCD renormalization scale $\mu$ and the NRQCD factorization scale $\mu_f$.

\item[(2)]
Due to a large cancellation at ${\cal O}(\alpha_s^3)$ between   the two nonconvergent matching coefficients, the  matching coefficient ratio is convergent up to N$^3$LO. Then by the approximation  in eq.~\eqref{ratiof}, we obtain the convergent decay constant ratio $f_{B_c^*}^{t,i0}/f_{B_c^*}^{t5,ij}$ up to N$^3$LO. Note that each physical decay constant is also convergent  (see  ref.~\cite{Tao:2023mtw}).

\item[(3)]
The N$^3$LO QCD correction to the ratio of the matching coefficients (decay constants) is almost independent of both $\mu_f$ and $\mu$, which verifies the correctness of our calculation for the decay constant ratio based on eq.~\eqref{ratiof} (also see related discussion in ref.~\cite{Tao:2023mtw}).

\item[(4)]
From the tables~\ref{tab:Cjnum} and \ref{tab:CivCjnum}, we also see the uncertainties of the matching coefficients and the matching coefficient (decay constant) ratio arising from the errors in the heavy quark masses $m_b$ and $m_c$ are relatively small compared to those resulting from the errors in $\mu_f$ and $\mu$ (also see ref.~\cite{Tao:2022qxa}).

\item[(5)]
For the $B_c^*$ decay constants involving different heavy flavor-changing currents, we    predict  $f_{B_c^*}^{v,i}=f_{B_c^*}^{t,i0}>f_{B_c^*}^{t5,ij}$.
\end{enumerate}


\section{Summary~\label{Summary}}

In this paper, we elaborate on the three-loop calculations of the NRQCD current renormalization constants (and corresponding anomalous dimensions), matching coefficients, (the ratio of) decay constants for the  heavy flavor-changing spatial-temporal tensor $(t,i0)$ current  and  spatial-spatial axial-tensor   $(t5,ij)$ current coupled to the $S$-wave vector $c\bar{b}$ meson $B_c^*$   within the NRQCD framework.
Although  the matching coefficients for both $(t,i0)$ and $(t5,ij)$ currents are nonconvergent, we can obtain the convergent ratio of  $B_c^*$ decay constants between $(t,i0)$ and $(t5,ij)$ currents up to N$^3$LO.
Our prediction for (the ratio of) $B_c^*$ decay constants involving (axial-)tensor currents, along with the experiment,
is useful to determine the fundamental parameters in particle physics and  is also of interest in beyond the Standard Model studies.

As a byproduct,  we obtain the three-loop finite term   for the ratio of QCD heavy flavor-changing tensor current renormalization constant in the $\mathrm{{OS}}$ scheme to that in the $\mathrm{\overline{MS}}$ scheme, which is a key ingredient to obtain  matching coefficients for  various heavy flavor-changing (axial-)tensor currents coupled to the $S$-wave and $P$-wave $c\bar{b}$ mesons.
And the study for  $P$-wave $c\bar{b}$ mesons is underway.

\hspace{2cm}

\noindent {\bf Acknowledgments:} 
We thank   A. Onishchenko and J. H. Piclum  for many helpful discussions.

\hspace{2cm}

%
%

\bibliographystyle{JHEP}
\bibliography{refs}

\end{document}